%% file: paper.tex
\documentclass[12pt]{article}

\usepackage{amsfonts}
\usepackage{amsmath}
\usepackage{amssymb}
\usepackage{calc}
\usepackage{fancyhdr}
\usepackage{float}
\usepackage{graphicx}
\usepackage{harvard}
\usepackage[margin=.70in]{geometry}
\usepackage{latexsym}
\usepackage{lscape}
\usepackage{multicol}
\usepackage{rotating}
\usepackage{t1enc}
\usepackage{theorem}

\numberwithin{equation}{section}
\renewcommand{\baselinestretch}{1.3}
\renewcommand{\cite}{\citeasnoun}

\newtheorem{theorem}{Theorem}

\newtheorem{corollary}{Corollary}

\newtheorem{definition}{Definition}

\newtheorem{lemma}{Lemma}

\newtheorem{remark}{Remark}

\begin{document}
\title{Realized range-based estimation of integrated variance\thanks{This paper was previously circulated under the title ``Asymptotic Theory for Range-Based Estimation of Integrated Variance of a Continuous Semi-Martingale.'' We thank Francis X. Diebold, Fulvio Corsi, Morten Ø. Nielsen, Peter R. Hansen, Roel C. A. Oomen, Svend E. Graversen, Tim Bollerslev, as well as conference and seminar participants at the 3$^{\text{rd}}$ Nordic Econometric Meeting in Helsinki, the FRU conference in Copenhagen, the Madrid meeting of the ``Microstructure of Financial Markets in Europe'' (MicFinMa) network, at Stanford University and Rady School of Management, UCSD, for helpful comments and suggestions. Special thanks go to Allan G. Timmermann, Asger Lunde, Holger Dette, Neil Shephard, two anonymous referees and the co-editor for providing insightful comments on earlier drafts. The second author is also grateful for financial assistance from the Deutsche Forschungsgemeinschaft through SFB 475 ``Reduction of Complexity in Multivariate Data Structures'' and funding from MicFinMa to support a six-month research visit at Aarhus School of Business. All algorithms for the paper were written in the Ox programming language, due to Doornik (2002). The usual disclaimer applies.}}
\author{Kim Christensen\thanks{Aarhus School of Business, Dept. of Marketing and Statistics, Haslegaardsvej 10, 8210 Aarhus V, Denmark. Phone: (+45) 89 48 63 74, fax: (+45) 86 15 39 88, e-mail: \texttt{kic@asb.dk}.}
\and
Mark Podolskij\thanks{Ruhr University of Bochum, Dept. of Probability and Statistics, Universit\"{a}tstrasse 150, 44780 Bochum, Germany. Phone: (+49) 234 32 28330, fax: (+49) 234 32 14559, e-mail: \texttt{podolski@cityweb.de}.}}

\date{November 7, 2006}

\maketitle

\begin{abstract}
We provide a set of probabilistic laws for estimating the quadratic variation of continuous semimartingales with realized range-based variance - a statistic that replaces every squared return of realized variance with a normalized squared range. If the entire sample path of the process is available, and under a set of weak conditions, our statistic is consistent and has a mixed Gaussian limit, whose precision is five times greater than that of realized variance. In practice, of course, inference is drawn from discrete data and true ranges are unobserved, leading to downward bias. We solve this problem to get a consistent, mixed normal estimator, irrespective of non-trading effects. This estimator has varying degrees of efficiency over realized variance, depending on how many observations that are used to construct the high-low. The methodology is applied to TAQ data and compared with realized variance. Our findings suggest that the empirical path of quadratic variation is also estimated better with the realized range-based variance.


\end{abstract}

\thispagestyle{empty}
\newpage

\section{Introduction}
\setcounter{page}{1} \renewcommand{\baselinestretch}{1.7}
\normalsize The volatility of asset prices is a key ingredient in
several areas of financial economics. Not long ago, academic studies
routinely used constant volatility models (e.g.,
\cite{black-scholes:73a}), despite empirical evidence in the data
suggesting that the conditional variance is both time-varying and
highly persistent. These facts were uncovered by the development and
application of parametric models, such as ARCH (see, e.g.,
\cite{bollerslev-engle-nelson:94a}), through stochastic volatility
models (e.g., \cite{ghysels-harvey-renault:96a}), and more recently
non-parametric methods based on high-frequency data, the most
conspicuous idea being \textit{realized variance} ($RV$), see, e.g.,
\cite{andersen-bollerslev-diebold-labys:01a} or
\cite{barndorff-nielsen-shephard:02a}; henceforth ABDL and BN-S.

$RV$ is the sum of squared returns over non-overlapping intervals
within a sampling period. Given weak regularity conditions, $RV$
converges in probability to the \textit{quadratic variation} ($QV$)
of all semimartingales as the sampling frequency tends to infinity.

In practice, the consistency of $RV$ breaks down as data limitations
prevent the sampling frequency from rising without bound. Most
notably, market microstructure noise contaminates high-frequency
asset prices. This invalidates the asymptotic theory, and $RV$ is
known to be inconsistent in the presence of noise (e.g.,
\cite[2006]{bandi-russell:05a} or \cite{hansen-lunde:06a}).
Therefore, it is common in applied work to construct $RV$ at a
moderate frequency, where the impact of noise is small enough to be
ignored, but this leads to loss of information. Though current
research seeks to make $RV$ robust against microstructure noise
(e.g., \cite{zhang-mykland-ait-sahalia:05a} or
\cite{barndorff-nielsen-hansen-lunde-shephard:08a}), the most
accurate estimator of $QV$ remains unknown. Set against this
backdrop, we suggest the \textit{realized range-based variance}
($RRV$).

Range-based estimation of volatility (developed in, e.g.,
\cite{feller:51a}, \cite{garman-klass:80a}, \cite{parkinson:80a},
\cite{rogers-satchell:91a}, \cite{kunitomo:92a}, and
\cite{alizadeh-brandt-diebold:02a}) reveals more information than
returns sampled at fixed intervals, because the extremes are formed
from the entire price process. The daily squared range, for example,
is about five times more efficient at estimating the scale of
Brownian motion than the daily squared return. But, as noted in
\cite{andersen-bollerslev:98a}, the accuracy of the high-low
estimator is only around that afforded by $RV$ based on two- or
three-hour returns, and the range has largely been neglected in the
recent literature.

Intraday range-based estimation of volatility, however, has the
potential of achieving smaller sampling errors than a sparsely
sampled $RV$, because we can replace every squared return of $RV$
with a squared range and extract most of the information about
volatility contained in the intermediate data points. No prior
studies have explored the properties of such an estimator. Indeed,
it is not clear what to expect from sampling, properly transformed,
high-frequency ranges. Extrapolating from the daily interval would
suggest that hourly ranges, say, achieve the accuracy of $RV$ based
on five- or ten-minute returns, but the comparison is more
complicated as each intraday range is constructed from less data.

We propose to sample and sum intraday price ranges to construct more
efficient estimates of $QV$. Our contributions are four-fold. First,
we develop a non-parametric method for measuring $QV$ with $RRV$.
Second, and unlike the existing time-invariant theory for the
high-low, we deal with estimation of time-varying volatility, when
the driving terms of the price process are (possibly) continuously
evolving random functions. Third, we derive a set of probabilistic
laws for sampling intraday high-lows. Fourth, we remove the problems
with downward bias reported in the previous range-based literature.
The new estimator is defined as:
\begin{equation}
RRV_{m}^{ \Delta} = \frac{1}{ \lambda_{2, m}} \sum_{i = 1}^{n}
s_{p_{i \Delta, \Delta}, m}^{2},
\end{equation}
where $s_{p_{i \Delta, \Delta}, m} = \max_{0 \leq s,t \leq m}
\left\{ p_{ \left(i - 1 \right)/n + t/mn} - p_{\left( i - 1
\right)/n + s/mn} \right\}$ is the observed range of a price process
$p$ over the interval $\left[ \left( i - 1 \right) / n,  i / n
\right]$, $i = 1, \ldots, n$. $m$ is the number of high-frequency
returns used to construct $s_{p_{i \Delta, \Delta}, m}$ and
$\lambda_{2, m}$ is a constant. We prove that $RRV_{m}^{ \Delta}$ is
consistent for the \textit{integrated variance} ($IV$) and that
$\sqrt{n} \left( RRV_{m}^{ \Delta} - IV \right)$ has a mixed
Gaussian limit with a variance that can be much smaller relative to
$RV$.

The paper is structured as follows. In the next section, we unfold
the necessary diffusion theory, present various ways of measuring
volatility and advance our methodological contribution by suggesting
$RRV$ and a version thereof that handles non-trading effects. Under
mild conditions, we prove consistency for the estimation method and
derive a mixed Gaussian central limit theorem (CLT). Section 3
illustrates the approach through Monte Carlo analysis to uncover the
finite sample properties, and we present some empirical results in
section 4. Rounding up, section 5 offers conclusions and sketches
several directions for future research.

\section{A semimartingale framework}
In this section, we propose a new method for consistently estimating
\textit{quadratic variation} ($QV$) based on the price range. The
theory is developed for the log-price of a univariate asset evolving
in continuous time over some interval, say $p = \left( p_{t}
\right)_{t \geq 0}$. $p$ is defined on a filtered probability space
$\bigl( \Omega, \mathcal{F}, \left( \mathcal{F}_{t} \right)_{t \geq
0}, \mathbb{P} \bigr)$ and adapted to the filtration $\left(
\mathcal{F}_{t} \right)_{t \geq 0}$, i.e. a collection of
$\sigma$-fields with $\mathcal{F}_{u} \subseteq \mathcal{F}_{t}
\subseteq \mathcal{F}$ for all $u \leq t < \infty$.

The basic building block is that $p$ constitutes a continuous sample
path semimartingale.\footnote{We adopt the continuity assumption as
a starting point only. In subsequent work, we have been analyzing
the properties of our estimator, when $p$ exhibits jumps (see
\cite{christensen-podolskij:12a}).} Hence, we write the time $t$
log-price in the generic form:
\begin{equation}
\label{PriceProcess} p_{t} = p_{0} + \int_{0}^{t} \mu_{u} \text{d}u
+ \int_{0}^{t} \sigma_{u} \text{d}W_{u}, \quad \text{for } t \geq 0,
\end{equation}
where $\mu = \left( \mu_{t} \right)_{t \geq 0}$ (the drift) is
locally bounded and predictable, $\sigma = \left( \sigma_{t}
\right)_{t \geq 0}$ (the volatility) is c\`{a}dl\`{a}g, and $W =
\left( W_{t} \right)_{t \geq 0}$ is a standard Brownian motion.

Much work in financial econometrics is cast within this setting
(see, e.g., \cite{andersen-bollerslev-diebold:10a} or BN-S (2007)
for reviews and references). Except for the continuity of the local
martingale, we impose little structure on the model. In fact, for
semimartingales with a continuous martingale component as above, the
form $\bigl( \int_{0}^{t} \mu_{u} \text{d}u \bigr)_{t \geq 0}$ is
implicit, when the drift term is predictable (in the absence of
arbitrage).\footnote{Moreover, all continuous local martingales,
whose $QV$ (to be defined in a moment) is absolutely continuous, has
the martingale representation of the second term in Equation
\eqref{PriceProcess}, e.g., \cite{doob:53a}. We refer to BN-S (2004,
footnote 6) for further details.}

The objective is to estimate a suitable measure of the return
variation over a subinterval $\left[ a, b \right] \subseteq \left[
0, \infty \right)$, labeled the sampling period or measurement
horizon. We assume $\left[ a, b \right] = \left[ 0, 1 \right]$; this
will be thought of as representing a trading day, but the choice is
arbitrary and can serve as a normalization. At any two sampling
times $t_{i - 1}$ and $t_{i}$, with $0 \leq t_{i - 1} \leq t_{i}
\leq 1$, the intraday return over $\left[ t_{i - 1}, t_{i} \right]$
is denoted by:
\begin{equation}
r_{t_{i}, \Delta_{i}} = p_{t_{i}} - p_{t_{i - 1}},
\end{equation}
where $\Delta_{i} = t_{i} - t_{i - 1}$.

From the theory of stochastic integration, it is well-known that
$QV$ is a natural measure of sample path variability for the class
of semimartingales. $QV$ is defined by:
\begin{equation}
\label{QV} \left\langle \, p \, \right\rangle = \underset{n \to
\infty}{ \text{p-} \negmedspace \lim} \sum_{i = 1}^{n} r_{t_{i},
\Delta_{i}}^{2},
\end{equation}
for any sequence of partitions, $0 = t_{0} < t_{1} < \ldots < t_{n}
= 1$, such that $\max_{1 \leq i \leq n} \left\{ \Delta_{i} \right\}
\to 0$ as $n \to \infty$ (e.g., \cite{protter:04a}).

In our framework, $QV$ is entirely induced by innovations to the
local martingale and coincides with the \textit{integrated variance}
($IV$), which is the object of interest:
\begin{equation}
\label{IV} IV = \int_{0}^{1} \sigma_{u}^{2} \text{d}u.
\end{equation}
$IV$ is central to financial economics, whether in asset and
derivatives pricing, portfolio selection or risk management (e.g.,
\cite{andersen-bollerslev-diebold:10a}). The econometric problem is
that $IV$ is latent, which complicates the empirical estimation of
this quantity. We briefly review the literature on existing methods
for measuring $IV$, before suggesting a new approach.

\subsection{Return-based estimation of integrated variance}
Not long ago, the daily squared return was employed as a
non-parametric estimator of $IV$. With the advent of high-frequency
data, however, more recent work has computed \textit{realized
variance} ($RV$), which is the sum of squared intraday returns
sampled over non-overlapping intervals (see, e.g., ABDL (2001) or
BN-S (2002)). More formally, consider an equidistant partition $0 =
t_{0} < t_{1} < \dots < t_{n} = 1$, where $t_{i} = i /
n$.\footnote{Though an irregular partition of the sampling period
suffices for consistency, it is standard to compute an equidistant
time series of intraday returns by various approaches, such as
linear interpolation in, e.g., \cite{andersen-bollerslev:97b} or the
previous-tick method suggested in
\cite{wasserfallen-zimmermann:85a}. A side-effect of linear
interpolation is that $RV^{ \Delta} \overset{p}{ \to} 0$ as $n \to
\infty$, because the interpolated process is of continuous bounded
variation, see \cite[Lemma 1]{hansen-lunde:06a}. Intuitively, a
straight line is the minimum variance path between two points.
\cite{oomen:05a} characterizes $RV$ under alternative sampling
schemes.} Then, adopting the notation of \cite{hansen-lunde:05a}, we
define $RV$ at sampling frequency $n$ by setting:
\begin{equation}
\label{Rv} RV^{ \Delta} = \sum_{i = 1}^{n} r_{i \Delta, \Delta}^{2}.
\end{equation}
$RV$ builds directly on the theory of $QV$. From Equation \eqref{QV}
and \eqref{IV}, it follows that
\begin{equation}
RV^{ \Delta} \overset{p}{ \to} IV,
\end{equation}
as $n \to \infty$.

BN-S (2002) derived a distribution theory for $RV^{ \Delta}$ in
relation to $IV$. The law of the scaled difference between $RV^{
\Delta}$ and $IV$ has a mixed Gaussian limit,
\begin{equation}
\label{CiDRv} \sqrt{n} \left( RV^{ \Delta} - IV \right) \overset{d}{
\to} MN \negmedspace \left( 0, 2 IQ \right),
\end{equation}
where
\begin{equation}
IQ = \int_{0}^{1} \sigma_{u}^{4} \text{d}u,
\end{equation}
is the \textit{integrated quarticity} ($IQ$). Thus, the size of the
error bounds for $RV^{ \Delta}$ is positively related to $\sigma$,
so $RV$ is a less precise estimator of $IV$ when $\sigma$ is high.
BN-S (2002) also derived a feasible CLT, where all quantities except
$IV$ can be computed directly from the data. This was done by simply
replacing $IQ$ by a consistent estimator, such as \textit{realized
quarticity} ($RQ$):
\begin{equation}
RQ^{ \Delta} = \frac{n}{3} \sum_{i = 1}^{n} r_{i \Delta,
\Delta}^{4},
\end{equation}
making it possible to construct confidence bands for $RV^{ \Delta}$
to measure the size of the estimation error involved with finite
sampling.

\subsection{Range-based estimation of integrated variance}
The choice of volatility proxy is not obvious in practice, since
microstructure bias affects $RV$ if $n$ is too large. With noisy
prices, $RV$ is both biased and inconsistent, see, e.g.,
\cite{zhou:96a}, \cite[2006]{bandi-russell:05a}, or
\cite{hansen-lunde:06a}.\footnote{With IID noise, for instance, $RV$
diverges to infinity, i.e. $RV^{ \Delta} \overset{p}{ \to} \infty$
as $n \to \infty$.} Previous studies have recognized this by
developing bias reducing techniques (e.g., pre-whitening of the
high-frequency return series with moving average or autoregressive
filters as in \cite{andersen-bollerslev-diebold-ebens:01a} and
\cite{bollen-inder:02a}, or kernel-based estimation as in
\cite{zhou:96a} and \cite{hansen-lunde:06a}).
\cite{zhang-mykland-ait-sahalia:05a} also suggest a subsample
estimator that is robust to noise in some situations. In empirical
work, the benefits of more frequent sampling is traded off against
the damage caused by cumulating noise, and - using various criteria
to pick the optimal sampling frequency - the result is often
sampling at a moderate frequency, e.g., every 5-, 10-, or
30-minutes, whereby data are discarded.

This pitfall of $RV$ motivates our choice of another proxy with a
long history in finance: the price range or high-low. Using the
terminology from above, we define the intraday range at sampling
times $t_{i - 1}$ and $t_{i}$ as:
\begin{equation}
s_{p_{t_{i}, \Delta_{i}}} = \underset{t_{i - 1} \leq s,t \leq t_{i}}
{\sup \left\{ p_{t} - p_{s} \right\}}.
\end{equation}
The subscript $p$ indicates that we use the range of the price
process. Below, we also need the range of a standard Brownian motion
over $\left[ t_{i - 1}, t_{i} \right]$, which is denoted by:
\begin{equation}
s_{W_{t_{i}, \Delta_{i}}} = \underset{t_{i - 1} \leq s,t \leq t_{i}}
{\sup \left\{ W_{t} - W_{s} \right\}}.
\end{equation}

\subsubsection{The distribution of the range}
\begin{figure}
    \begin{centering}
        \includegraphics[height=10.00cm,width=\linewidth]
            {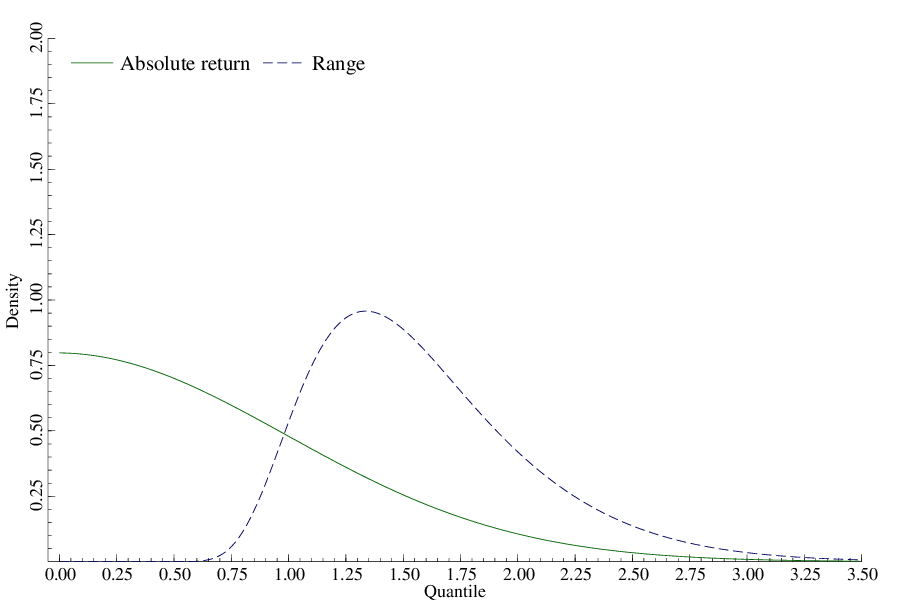}\\
    \end{centering}
    \caption{The distribution of the absolute return and range of a
    standard Brownian motion over an interval of unit length.}
    \label{AsPdf.eps}
\end{figure}

The foundations of the range go back to \cite{feller:51a}, who found
its distribution by using the theory of Brownian
motion.\footnote{There are two types of range-based volatility
estimators: The first relies purely on the high-low, while the
second combines the high-low with the open-close, e.g.,
\cite{garman-klass:80a} or \cite{rogers-satchell:91a}. Throughout,
we only consider the high-low estimator.} According to his work, the
density of $s_{W_{t_{i}, \Delta_{i}}}$ is given by:
\begin{equation}
\label{RbPPdf} \text{f} \left( x \right) = 8 \sum_{j = 1}^{ \infty}
\left( -1 \right)^{j - 1} \frac{j^{2}}{ \sqrt{ \Delta_{i}}} \phi
\left( \frac{jx}{ \sqrt{ \Delta_{i}}} \right), \quad \text{for } x
> 0,
\end{equation}
with $\phi \left( y \right) = \exp \left(- y^{2}/2 \right) / \sqrt{2
\pi}$. The infinite series is evaluated by a suitable truncation. In
Figure \ref{AsPdf.eps}, we plot the density function of
$s_{W_{t_{i}, \Delta_{i}}}$ by taking $t_{i} = \Delta_{i} = 1$ (We
use the shorthand notation $s_{W}$ for this random variable in the
rest of the paper).

\begin{center}
[ INSERT FIGURE \ref{AsPdf.eps} ABOUT HERE ]
\end{center}

The figure also displays the distribution of the absolute return. By
comparing these proxies, it is suggestive that the efficiency of the
range is higher, or in other words that its variance vis-\`{a}-vis
the return is lower.

\cite{parkinson:80a} used Feller's insights to derive the moment
generating function of the range of a scaled Brownian motion, $p_{t}
= \sigma W_{t}$.\footnote{Note, $\sigma$ does double-duty;
representing either the process $\sigma = \left( \sigma_{t}
\right)_{t \geq 0}$ or a constant diffusion parameter $\sigma_{t} =
\sigma$. The meaning is clear from the context.} For the $r$th
moment:
\begin{equation}
\label{RbPu} \mathbb{E} \left[ s_{p_{t_{i}, \Delta_{i}}}^{r} \right]
= \lambda_{r} \Delta_{i}^{r / 2} \sigma^{r}, \quad \text{for } r
\geq 1,
\end{equation}
where $\lambda_{r} = \mathbb{E} \left[ s_{W}^{r}
\right]$.\footnote{The explicit formula for $\lambda_{r}$ is:
$\lambda_{r} = \frac{4}{ \sqrt{ \pi}} ( 1 - \frac{4}{2^{r}} ) 2^{r /
2} \Gamma ( \frac{r + 1}{2} ) \zeta \left( r - 1 \right)$, for $r
\geq 1$; where $\Gamma \left( x \right)$ and $\zeta \left( x
\right)$ denote the Gamma and Riemann's zeta function,
respectively.}

Arguably, a process without drift and constant $\sigma$ is
irrelevant from an empirical point of view. An overwhelming amount
of research indicates that $\sigma$ is time-varying, see, e.g.,
\cite{ghysels-harvey-renault:96a}. Nonetheless, to our knowledge
there exists little theory about range-based estimation of $IV$ in
the presence of a continually evolving diffusion
parameter.\footnote{A notable exception is
\cite{gallant-hsu-tauchen:99a}, who estimate two-factor stochastic
volatility models in a general continuous time framework. They
derive the density function of the range in this setting, but do not
otherwise explore its theoretical properties.} Previous work
accounts for (randomly) changing volatility by holding $\sigma_{t}$
fixed within the trading day, while allowing for (stochastic) shifts
between them (e.g., \cite{alizadeh-brandt-diebold:02a}). Still,
there are strong intraday movements in $\sigma_{t}$ (e.g.,
\cite{andersen-bollerslev:97b}).

A major objective of this paper is therefore to extend the
theoretical domain of the extreme value method to a more general
class of stochastic processes. Contrary to extant research, we
develop a statistical framework for the Brownian semimartingale in
Equation \eqref{PriceProcess}, featuring less restrictive dynamics
for $\mu$ and $\sigma$.

\subsubsection{A realized range-based estimator}
As stated earlier, the (transformed) daily range is less efficient
than $RV$ for moderate values of $n$; two- or three-hour returns
suffice. But with tick-by-tick data at hand, we can construct more
precise range-based estimates of $IV$ by sampling high-lows within
the trading day. Curiously, a rigorous analysis of intraday ranges
has been missing in the volatility literature.\footnote{In an
independent, concurrent paper, \cite{martens-dijk:07a} have studied
$RRV$ for homoscedastic diffusions, but they do not derive a general
asymptotic theory.}

Accordingly, consider again the equidistant partition with $t_{i} =
i / n$, for $i = 1, \ldots, n$.\footnote{We use equidistant
estimation to ease notation. All our results generalize to an
irregular subdivision of the sampling period, so long as $\max_{1
\leq i \leq n} \left\{ \Delta_{i} \right\} \to 0$ as $n \to \infty$,
although the conditional variance in the CLT is modified slightly,
as spelled out below.} We then propose a \textit{realized
range-based variance} ($RRV$) estimator of $IV$, which - at sampling
frequency $n$ - is defined as:
\begin{equation}
\label{RbP} RRV^{ \Delta} = \frac{1}{ \lambda_{2}} \sum_{i = 1}^{n}
s_{p_{i \Delta, \Delta}}^{2}.
\end{equation}
$RRV^{ \Delta}$ has two advantages over the previous return- and
range-based methods suggested in the literature on volatility
estimation. First, $RRV^{ \Delta}$ inspects all data points
(regardless of $n$), whereby we avoid neglecting information about
$IV$. Second, the efficiency of $RRV^{ \Delta}$ is several times
that of $RV^{ \Delta}$, leading to narrower confidence intervals for
$IV$ (see below).

\subsubsection{Properties of realized range-based variance}
The properties of $RRV^{ \Delta}$ are trivial for the scaled
Brownian motion, $p_{t} = \sigma W_{t}$. As the infill asymptotics
start operating by letting $n \to \infty$, we achieve an increasing
sequence of IID random variables, $\left\{ s_{p_{i \Delta, \Delta}}
\right\}_{i = 1, \dots, n}$. Suitably transformed to unbiased
measures of $\sigma^{2}$ using \eqref{RbPu}, the consistency of
$RRV^{ \Delta}$ follows from a standard law of large numbers by
averaging. To see this, note that $\mathbb{E} \left[ RRV^{ \Delta}
\right] = \sigma^{2}$ and $\text{var} \left[ RRV^{ \Delta} \right] =
\Lambda n^{-1} \sigma^{4}$ with:
\begin{equation}
\Lambda = \frac{ \lambda_{4} - \lambda_{2}^{2}}{ \lambda_{2}^{2}}
\simeq 0.4073.
\end{equation}
Hence, $RRV^{ \Delta} \overset{p}{ \to} \sigma^{2}$ as $n \to
\infty$. Also, for this process a stardard CLT implies that:
\begin{equation}
\sqrt{n} \left( RRV^{ \Delta} - \sigma^{2} \right) \overset{d}{ \to}
N \negmedspace \left( 0, \Lambda \sigma^{4} \right).
\end{equation}
If $\mu$ and $\sigma$ are stochastic, establishing the large sample
properties of $RRV^{ \Delta}$ is more involved, but nonetheless
possible. Overall, the basic idea extends to general Brownian
semimartingales, given some regularity on $\mu$ and $\sigma$, as we
next show.\footnote{Throughout the paper, proofs of the theorems are
presented in the Appendix.}

\begin{theorem}
\label{RbPCiP} Assume $p$ satisfies the continuous time stochastic
volatility model in Equation \eqref{PriceProcess}, where $\mu$ is
locally bounded and predictable, and $\sigma$ is c\`{a}dl\`{a}g.
Then, as $n \to \infty$,
\begin{equation}
RRV^{ \Delta} \overset{p}{ \to} IV.
\end{equation}
\end{theorem}
No knowledge about the dynamics of $\sigma$ is needed for Theorem
\ref{RbPCiP} to hold, except for weak technical conditions, so it
considerably extends the theory of range-based volatility
estimation. We allow for very general continuous time processes,
including, but not limited to, models with leverage, long-memory,
diurnal effects or jumps (in $\sigma$). This is certainly not true
in the previous range-based literature. Moreover, the theorem allows
for drift due to the fact that the variation induced by the expected
move in $p$, $\bigl( \int_{0}^{t} \mu_{u} \text{d}u \bigr)_{t \geq
0}$, is an order of magnitude lower than the variation induced by
the continuous local martingale; comprised by $\bigl( \int_{0}^{t}
\sigma_{u} \text{d}W_{u} \bigr)_{t \geq 0}$.

\subsubsection{Asymptotic distribution theory}
In empirical work, the consistency of $RRV^{ \Delta}$ becomes
unreliable due to microstructure noise, if $n$ is too large. Theorem
\ref{RbPCiP} does not indicate the precision of $RRV^{ \Delta}$ if
$n$ is fixed at a moderate level, and econometricians often compute
confidence bands as a guide to the error made from estimation in
finite samples. To strengthen the convergence in probability, we
next develop a distribution theory for $RRV^{ \Delta}$.

The above weak assumptions on $\sigma$ are too general to prove a
CLT, and we need slightly stronger conditions:\\[-0.25cm]

\noindent \textbf{Assumption (V)}: \textit{$\sigma$ does not vanish
(V$_{1}$) and satisfies:
\begin{equation}
\tag{ \textbf{V}$_{ \mathbf{2}}$} \sigma_{t} = \sigma_{0} +
\int_{0}^{t} \mu_{u}^{ \prime} \text{\upshape{d}}u + \int_{0}^{t}
\sigma_{u}^{ \prime} \text{\upshape{d}}W_{u} + \int_{0}^{t} v_{u}^{
\prime} \text{\upshape{d}}B_{u}^{ \prime}, \quad \text{\upshape{for
}} t \geq 0,
\end{equation}
where $\mu^{ \prime} = \left( \mu_{t}^{ \prime} \right)_{t \geq 0}$,
$\sigma^{ \prime} = \left( \sigma_{t}^{ \prime} \right)_{t \geq 0}$
and $v^{ \prime} = \left( v_{t}^{ \prime} \right)_{t \geq 0}$ are
c\`{a}dl\`{a}g, with $\mu^{ \prime}$ also being locally bounded and
predictable, and $B^{ \prime} = \left( B_{t}^{ \prime} \right)_{t
\geq 0}$ is a Brownian motion independent of $W$.}\\[-0.25cm]

We prove our result by invoking stable convergence in law. This is
standard in the $RV$ literature. But to avoid any confusion about
our terminology, we present the definition.
\begin{definition}
\label{StableC} A sequence of random variables, $\left( X_{n}
\right)_{n \in \mathbb{N}}$, \textup{converges stably in law with
limit $X$}, defined on an appropriate extension of $\bigl( \Omega,
\mathcal{F}, \left( \mathcal{F}_{t} \right)_{t \geq 0}, \mathbb{P}
\bigr)$, if and only if for every $\mathcal{F}$-measurable, bounded
random variable $Y$ and any bounded, continuous function $g$, the
convergence $\lim_{n \to \infty} \mathbb{E} \left[Y g \left(X_{n}
\right) \right] = \mathbb{E} \left[ Y g \left(X \right) \right]$
holds.
\end{definition}
We use the symbol $X_{n} \overset{d_{s}}{ \to} X$ to denote stable
convergence. Note that this implies weak convergence, which may be
equivalently defined by taking $Y = 1$ (see, e.g., \cite{renyi:63a}
or \cite{aldous-eagleson:78a} for more details).

We now state the main result, which is a (non-standard) CLT.
\begin{theorem}
\label{RbPCiD} Assume that the conditions of Theorem \ref{RbPCiP}
hold and assumption \textbf{(V)} is satisfied. Then it holds that,
as $n \to \infty$,
\begin{equation}
\label{EqRbPCiD} \sqrt{n} \left( RRV^{ \Delta} - IV \right)
\overset{d_{s}}{ \to} \sqrt{ \Lambda} \int_{0}^{1} \sigma_{u}^{2}
\text{\textup{d}}B_{u},
\end{equation}
where $B = \left( B_{t} \right)_{t \geq 0}$ is a standard Brownian
motion, independent from $\mathcal{F}$ (written
$B~\raisebox{-0.3ex}{\rotatebox{90}{$ \models$}}~\mathcal{F}$).
\end{theorem}
A critical feature of this theorem is that the left-hand side
converges to a stochastic integral with respect to $B$, which is
independent of the driving term $\sigma$. This implies $\sqrt{n}
\left( RRV^{ \Delta} - IV \right)$ has a mixed normal limit, with
$\sigma$ governing the mixture.\footnote{Earlier drafts of this
paper had a non-mixed Gaussian CLT and the stronger conditions, $\mu
= 0$ and $\sigma$ is H\"{o}lder continuous of order $\gamma > 1/2$,
i.e. $\sigma_{t} - \sigma_{s} = \text{O}_{p} \left( \mid t - s
\mid^{ \gamma} \right)$ for $t \to s$. We have substantially
weakened these restrictions and also proved the mixed Gaussian CLT.
Svend E. Graversen was helpful in pointing our attention to a result
that enabled us to remove these assumptions (see Lemma \ref{Unique}
in the Appendix).} In general, this introduces heavier tails in the
unconditional distribution of $RRV^{ \Delta}$ than for Gaussian
random variables. To summarize:
\begin{equation}
\label{AmnRbP} \sqrt{n} \left( RRV^{ \Delta} - IV \right)
\overset{d}{ \to} MN \negmedspace \, \left( 0, \Lambda IQ \right).
\end{equation}
\begin{remark}
The $\Lambda$ scalar in front of $IQ$ in Equation \eqref{AmnRbP} is
roughly 0.4. In contrast, the number appearing in the CLT for $RV^{
\Delta}$ is $2$.
\end{remark}
Hence, the sampling errors of $RRV^{ \Delta}$ are about one-fifth of
those based on $RV^{ \Delta}$. This is not surprising: $RRV^{
\Delta}$ uses all the data, whereas $RV^{ \Delta}$ is based on
high-frequency returns sampled at fixed points in time. As, for the
moment, $p$ is assumed fully observed, $RV^{ \Delta}$ is neglecting
a lot of information.

$IQ$ on the right-hand side in \eqref{AmnRbP} is infeasible, i.e. it
cannot be computed directly from the data. We can estimate it with
the \textit{realized range-based quarticity} ($RRQ$):
\begin{equation}
RRQ^{ \Delta} = \frac{n}{ \lambda_{4}} \sum_{i = 1}^{n} s_{p_{i
\Delta, \Delta}}^{4}.
\end{equation}
With techniques similar to the proof of Theorem \ref{RbPCiP}, we can
show that $RRQ^{ \Delta} \overset{p}{ \to} IQ$. Thus, by using the
properties of stable convergence (e.g., \cite{jacod:97a}), we get
the next corollary.
\begin{corollary}
\label{RbPAsN} Given the conditions of Theorem \ref{RbPCiD}, it
follows that:
\begin{equation}
\frac{ \sqrt{n} \left( RRV^{ \Delta} - IV \right)}{ \sqrt{ \Lambda
RRQ^{ \Delta}}} \overset{d}{ \to} N(0, 1).
\end{equation}
\end{corollary}
\begin{remark}
With irregular sampling schemes, the distributional result in
\eqref{AmnRbP} - and those in the next sections  - changes slightly
(the stochastic limit is unchanged). Set
\begin{align}
RRV^{ \Xi} &= \frac{1}{ \lambda_{2}} \sum_{i = 1}^{n} s_{p_{t_{i},
\Delta_{i}}}^{2}, \\
H^{ \Xi}_{n, u} &= n \sum_{i = 1}^{j : t_{j} \leq u} \left( t_{i} -
t_{i - 1} \right)^{2},
\end{align}
and assume that a pointwise limit $H^{ \Xi}_{u}$ of $H^{ \Xi}_{n,
u}$ exists and is continuously differentiable. Then, as $n \to
\infty$ such that $\max_{1 \leq i \leq n} \left\{ \Delta_{i}
\right\} \to 0$:
\begin{equation}
\label{AmnRbPXi} \sqrt{n} \left( RRV^{ \Xi} - IV \right)
\overset{d}{ \to} MN \negmedspace \left( 0, \Lambda \int_{0}^{1}
\frac{ \partial H^{ \Xi}_{u}}{ \partial u} \sigma_{u}^{4}
\text{\textup{d}}u \right).
\end{equation}
The derivative $\partial H^{ \Xi}_{u} / \partial u$ is small, when
sampling runs quickly. Hence, there are potential gains in having
more frequent observations when $\sigma$ is high.
\cite{hansen-lunde:06a} prove that such a sampling scheme minimizes
the asymptotic variance of $RV$. Obviously, for equidistant
subdivisions $H^{ \Xi}_{u} = u$, so the extra term drops out. The
theory is made feasible with
\begin{equation}
RRQ^{ \Xi} = \frac{n}{ \lambda_{4}} \sum_{i = 1}^{n} s_{p_{t_{i},
\Delta_{i}}}^{4} \overset{p}{ \to} \int_{0}^{1} \frac{ \partial H^{
\Xi}_{u}}{ \partial u} \sigma_{u}^{4} \text{\textup{d}}u.
\end{equation}
\end{remark}

\subsubsection{Discretely sampled high-frequency data}
In practice, we draw inference about $IV$ from a finite data sample
and cannot extract the true range, so the intraday high-low
statistic will be progressively more downward biased as $n$ gets
larger. Building on the simulation evidence of
\cite{garman-klass:80a}, \cite{rogers-satchell:91a} proposed a
technique for bias correcting the range that largely removed the
error from a numerical perspective.

Nonetheless, it is misleading to think about ranges as downward
biased. The source of the bias is $\lambda_{2}$, which is
constructed on the presumption that $p$ is fully observed.
Therefore, we will now develop an estimator that accounts for the
number of high-frequency data points used in forming the high-low,
in order to scale properly. To formalize this idea, additional
notation is required. Assume, without loss of generality, that $mn +
1$ equidistant observations of the price process are available,
giving $mn$ returns. These are split into $n$ intervals each with
$m$ innovations. We denote the observed range over the $i$th
interval by:
\begin{equation}
s_{p_{i \Delta, \Delta}, m} = \max_{0 \leq s,t \leq m} \left\{ p_{
\left(i - 1 \right)/n + t/mn} - p_{\left( i - 1 \right)/n + s/mn}
\right\}.
\end{equation}
Also, we let:
\begin{equation}
s_{W, m} = \max_{0 \leq s,t \leq m} \left\{ W_{t/m} - W_{s/m}
\right\},
\end{equation}
and then define a new realized range-based estimator by setting:
\begin{equation}
\label{RbPm} RRV_{m}^{ \Delta} = \frac{1}{ \lambda_{2, m}} \sum_{i =
1}^{n} s_{p_{i \Delta, \Delta}, m}^{2},
\end{equation}
where $\lambda_{r, m} = \mathbb{E} \bigl[ s_{W, m}^{r} \bigr]$.
$\lambda_{r, m}$ is the $r$th moment of the range of a standard
Brownian motion over a unit interval, when we only observe $m$
increments of the underlying continuous time process.

To our knowledge, there is no explicit formula for $\lambda_{r, m}$,
but it is easily simulated to any degree of accuracy. Figure
\ref{L(2,m).eps} details this for $r = 2$ and all values of $m$ that
integer divide 23,400.

\begin{center}
[ INSERT FIGURE \ref{L(2,m).eps} ABOUT HERE ]
\end{center}

Of course, $\lambda_{2, m} \to \lambda_{2}$ as $m \to \infty$, but
note also that $\lambda_{2, 1} = 1$, which defines $RV^{ \Delta}$.
The downward bias reported in simulation studies on the range-based
estimator is a consequence of the fact that $1 / \lambda_{2}$ was
applied in place of $1 / \lambda_{2, m}$, as the bias is in
one-to-one correspondence with the difference.

Having completed these preliminaries, we prove consistency and
asymptotic normality for the estimator in Equation \eqref{RbPm}.
Note that $m$ is not required to approach infinity for the CLT to
work; convergence to any natural number is sufficient.
\begin{theorem}
\label{RbPCiDm} Given the assumptions of Theorem \ref{RbPCiP}, as $n
\to \infty$,
\begin{equation}
RRV_{m}^{ \Delta} \overset{p} \to IV,
\end{equation}
where the convergence is uniform in $m$. Moreover, if \textbf{(V)}
holds and $m \to c \in \mathbb{N} \cup \left\{ \infty \right\}$:
\begin{equation}
\sqrt{n} \left( RRV_{m}^{ \Delta} - IV \right) \overset{d_{s}}{ \to}
\sqrt{\Lambda_{c}} \int_{0}^{1} \sigma_{u}^{2}
\text{\textup{d}}B_{u},
\end{equation}
where $\Lambda_{c} = \bigl( \lambda_{4, c} - \lambda_{2, c}^{2}
\bigr) / \lambda_{2, c}^{2}$ and
$B~\raisebox{-0.3ex}{\rotatebox{90}{$ \models$}}~\mathcal{F}$.
Finally,
\begin{equation}
\label{EqRbPCiDm} \frac{ \sqrt{n} \left( RRV_{m}^{ \Delta} - IV
\right)}{ \sqrt{ \Lambda_{m} RRQ_{m}^{ \Delta}}} \overset{d}{ \to}
N(0, 1),
\end{equation}
with $\Lambda_{m} = \left( \lambda_{4, m} - \lambda_{2, m}^{2}
\right) / \lambda_{2, m}^{2}$ and
\begin{equation}
RRQ_{m}^{ \Delta} = \frac{n}{ \lambda_{4, m}} \sum_{i = 1}^{n}
s_{p_{i \Delta, \Delta}, m}^{4}.
\end{equation}
\end{theorem}
\begin{remark}
Theorem \ref{RbPCiDm} provides a CLT for $RV^{ \Delta}$ with $m =
1$, as also derived in, e.g., BN-S (2002) or
\cite{barndorff-nielsen-graversen-jacod-podolskij-shephard:06a}.
\end{remark}

\begin{figure}
    \begin{centering}
        \includegraphics[height=10.00cm,width=\linewidth]
            {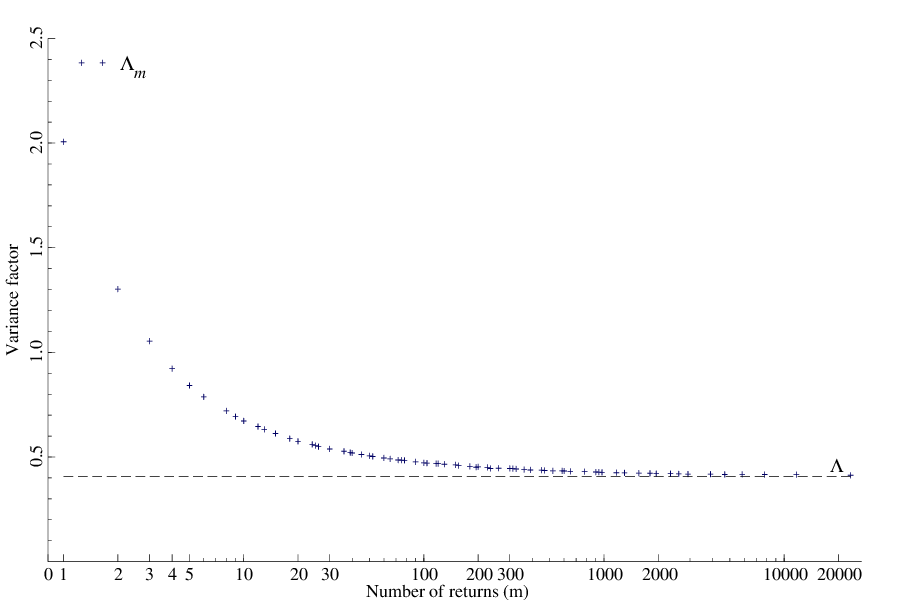}\\
    \end{centering}
    \caption{We plot $\Lambda_{m}$, the variance factors appearing
    in the CLT of $RRV_{m}^{ \Delta}$, which are estimated from a
    simulation with 1,000,000 repetitions. The dashed line is the
    asymptotic value.}
    \label{CLT.eps}
\end{figure}

\begin{center}
[ INSERT FIGURE \ref{CLT.eps} ABOUT HERE ]
\end{center}

To provide an impression of the efficiency of $RRV^{ \Delta}_{m}$,
Figure \ref{CLT.eps} depicts $\Lambda_{m}$ on the y-axis, as a
function of $m$ along the x-axis. The steep initial decline in
$\Lambda_{m}$ renders the advantage of $RRV^{ \Delta}_{m}$ large
compared to $RV^{ \Delta}$ even for moderate values of $m$. For $m =
10$, say, since the scalar appearing in front of $IQ$ in the CLT for
$RRV^{ \Delta}_{m}$ is about 0.7, the confidence intervals for $IV$
are much narrower. In our experience $m = 10$, or higher values, is
usually obtained for moderately liquid assets at empirically
relevant frequencies, such as 5-minute sampling.\footnote{Under
parametric assumptions and no microstructure noise, $RV$ is the
maximum likelihood estimator. Thus, our efficiency comparison should
be viewed as the potential reduction in variance that can be
achieved with $RRV_{m}^{ \Delta}$ when microstructure noise is
preventing $RV^{ \Delta}$ from being sampled at the maximum
frequency ($mn$) and $n$ is set at a moderate level where the impact
of noise is minimal.}

\section{Monte Carlo experiment}
To study the finite sample properties of $RRV^{ \Delta}_{m}$, this
section uses repeated samples from a stochastic volatility model. We
simulate the following system of stochastic differential equations:
\begin{equation}
\label{SimLn}
\begin{array}{r@{~}l}
\text{d}p_{t} &= \sigma_{t} \text{d}W_{t}, \\
\text{d} \ln \sigma_{t}^{2} &= \theta ( \omega - \ln \sigma_{t}^{2})
\text{d}t + \eta \text{d}B_{t},
\end{array}
\end{equation}
where $W$ and $B$ are independent Brownian motions, while $\left(
\theta, \omega, \eta \right)$ are parameters. Thus, the log-variance
of spot prices evolves as a mean reverting Ornstein-Uhlenbeck
process with mean $\omega$, mean reversion parameter $\theta$ and
volatility $\eta$ (see, e.g., \cite{gallant-hsu-tauchen:99a},
\cite{alizadeh-brandt-diebold:02a}, and
\cite{andersen-benzoni-lund:02a}). The vector $\left( \theta,
\omega, \eta \right) = \left( 0.032, -0.631, 0.115 \right)$ is taken
from \cite{andersen-benzoni-lund:02a}, who apply Efficient Method of
Moments (EMM) to calibrate numerous continuous time models.

Initial conditions are set at $p_{0} = 0$ and $\ln \sigma_{0}^{2} =
\omega$, and we generate $T = 1,000,000$ daily replications from
this model each with $mn$ returns, where $mn$ depends on the setting
(see below). Throughout, we continue to ignore the irregular spacing
of empirical high-frequency data and work with equidistant data.

\subsection{Simulation results}
The distributional result for $RRV^{ \Delta}_{m}$ is detailed by
setting $m = 10$. The reported results are not very sensitive to
specific choices of $m$, but in general higher values improve the
coverage rates of the asymptotic confidence bands. We simulate $n =
10$, $50$, $100$ for a total of $mn = 100$, $500$, $1000$ increments
each day, allowing us to show the convergence in distribution to the
standard normal for high-frequency sample sizes that resemble those
of moderately liquid assets.

\begin{figure}
    \begin{centering}
            \includegraphics[height=9.00cm,width=\linewidth]
                {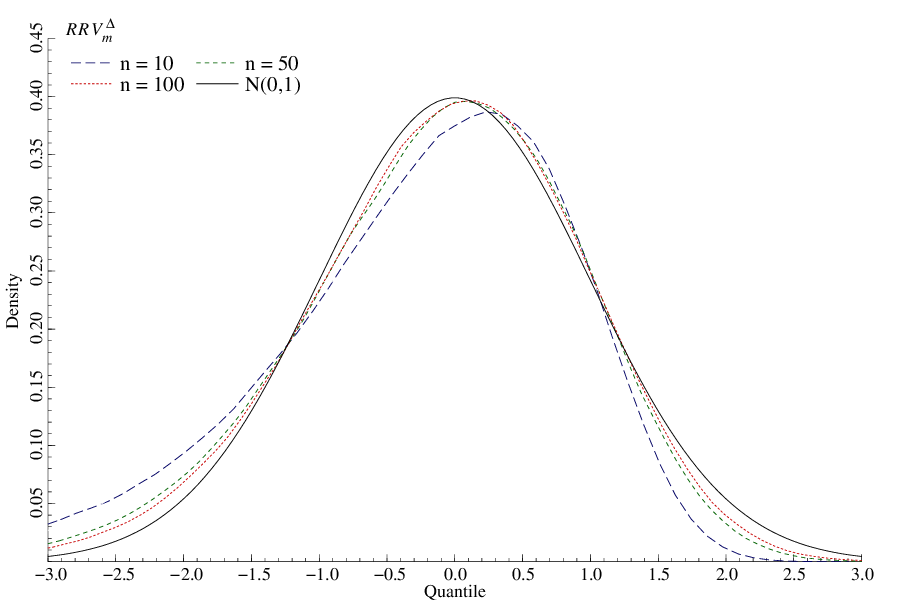}
            \includegraphics[height=9.00cm,width=\linewidth]
                {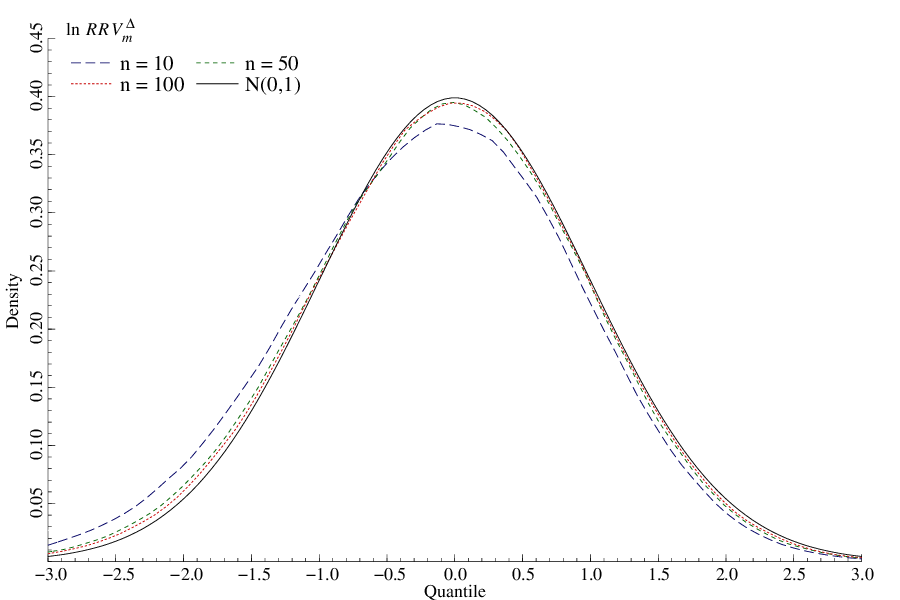}\\
    \end{centering}
    \caption{Asymptotic normality for the standardized realized
    range-based statistic in estimating $IV$. The figure plots
    kernel densities of the sampling errors of $RRV_{m}^{ \Delta}$
    for the small sample settings $n = 10$, $50$, $100$ and $m =
    10$. All plots are based on a simulation with 1,000,000
    repetitions from a log-normal diffusion for $\sigma$, as
    explained in the main text. The upper panel depicts t-statistics
    of the feasible CLT for $RRV_{m}^{ \Delta}$, while the lower
    panel is the corresponding log-based version. The solid line is
    the N(0,1) density.}
    \label{RbP-SvAsN-m=10.eps}
\end{figure}

\begin{center}
[ INSERT FIGURE \ref{RbP-SvAsN-m=10.eps} ABOUT HERE ]
\end{center}

Figure \ref{RbP-SvAsN-m=10.eps} (upper panel) graphs kernel
densities for the standardized errors of $RRV_{m}^{ \Delta}$; cf.
the ratio in Equation \eqref{EqRbPCiDm}. For $n = 10$, the
distribution is left-skewed with a poor approximation in both the
center and tail areas compared to the N(0,1) reference density. The
distortions are diminished by progressively increasing the sample.
With $n = 100$ the tails are tracked quite closely.

BN-S (2005) showed that log-based inference via standard
linearization methods improved the raw distribution theory for $RV^{
\Delta}$. They reported better finite sample behavior for the errors
of the log-transform than those extracted with the feasible version
of the CLT outlined in Equation \eqref{CiDRv}. The shape of the
actual densities for $RRV_{m}^{ \Delta}$ suggests that this also
applies to our setting. By the delta method, the log-version of the
CLT for $RRV_{m}^{ \Delta}$ takes the form:
\begin{equation}
\label{RbPCiDLn} \sqrt{n} \left( \ln RRV_{m}^{ \Delta} - \ln IV
\right) \overset{d}{ \to} MN \negmedspace \left( 0, \frac{
\Lambda_{c} IQ}{IV^{2}} \right).
\end{equation}
In the lower panel of Figure \ref{RbP-SvAsN-m=10.eps}, we plot the
density functions of the feasible log-based t-statistics. The
coverage probabilities of Equation \eqref{RbPCiDLn} are a much
better guide for small values of $n$, with $n = 100$ providing a
near perfect fit to the N(0,1) distribution. Hence, the results for
$RRV_{m}^{ \Delta}$ are consistent with the findings for $RV^{
\Delta}$.

This technique is also applicable to study other (differentiable)
functions of $RRV_{m}^{ \Delta}$. For convenience, we state the CLT
of a particularly useful transformation, obtained by taking square
roots:
\begin{equation}
\sqrt{n} \left( \sqrt{RRV^{ \Delta}_{m}} - \sqrt{IV} \right)
\overset{d}{ \to} MN \negmedspace \left( 0, \frac{ \Lambda_{c} IQ}{4
IV} \right).
\end{equation}

\section{Empirical application: General Motors}
We investigate the empirical properties of intraday ranges by
analyzing a major stock from the Dow Jones Industrial Average,
General Motors (GM).

High-frequency data were extracted from the TAQ database, which is a
recording of trades and quotes from the securities listed on New
York Stock Exchange (NYSE), American Stock Exchange (AMEX), and
National Association of Securities Dealers Automated Quotation
(NASDAQ). The sample period covers January 3, 2000 through December
31, 2004; a total of 1,255 trading days. We restrict attention to
NYSE updates and only report the results of the quotation data, for
which the midquote is used.\footnote{The analysis of transaction
data is available upon request.} All raw data were filtered for
irregularities (e.g., prices of zero, entries posted outside the
NYSE opening hours, or quotes with negative spreads), and a second
algorithm handled outliers in the price series.

\begin{center}
[ INSERT TABLE \ref{GM-DataC.tex} ABOUT HERE ]
\end{center}

The average number of data points after filtering is given in Table
\ref{GM-DataC.tex}. The column \#$r_{ \tau_{i}} \neq 0$, where $r_{
\tau_{i}} = p_{ \tau_{i}} - p_{ \tau_{i - 1}}$ and $\tau_{i}$ is the
arrival time of the $i$th tick, counts the number of price changes
relative to the previous posting. \#$\Delta r_{ \tau_{i}} \neq 0$
does the same for second differences, but after having removed
updates with $r_{ \tau_{i}} = 0$. These numbers are important to
calculate $\lambda_{2, m}$ and $\lambda_{4, m}$ that are required to
estimate $RRV_{m}^{ \Delta}$ and construct confidence bands.
Initially, we found $mn$ on the basis of all non-zero returns; i.e.
the \#$r_{ \tau_{i}} \neq 0$ numbers. This meant $mn$ was too high,
because of instantaneous reversals (e.g., bid-ask bounce behavior).
We assessed that a proper method to determine $mn$ was to only count
repeated reversals once. Thus, to compute $mn$ we use the \#$\Delta
r_{ \tau_{i}} \neq 0$ numbers.

The estimation of $RV^{ \Delta}$ and $RRV_{m}^{ \Delta}$ proceeds
with 5-minute sampling through the trading session starting
9:30{\scriptsize AM} EST until 4:00{\scriptsize PM} EST; i.e. by
setting $n = 78$ or $\Delta = 300$ seconds.\footnote{This choice was
guided by signature plots, i.e. sample averages of the estimators
across different sampling frequencies $n$. We found increasing signs
of microstructure noise by moving below the 5-minute frequency.} We
use the previous-tick method to compute returns for $RV^{ \Delta}$.
Note that since the empirical high-frequency data are irregularly
distributed, there are, in general, different values of $m$ in the
5-minute intervals. This does not cause any problems, however, for
the theory extends directly to this setting, provided we use the
individual values of $m$ in the estimation.

\begin{center}
[ INSERT TABLE \ref{GM-qu-Sample.tex} ABOUT HERE ]
\end{center}

Sample statistics for the resulting time series are printed in Table
\ref{GM-qu-Sample.tex}. $RV^{ \Delta}$ has a lower minimum and a
higher maximum than $RRV_{m}^{ \Delta}$, while its overall mean is
higher. Both kurtosis figures are consistent with a mixed Gaussian
limit. The variance of $RRV_{m}^{ \Delta}$ is only 58\% that of
$RV^{ \Delta}$. This is much lower, but as expected still somewhat
higher than predicted by the theory (relative to $RV^{ \Delta}$).
First off, here we are looking at a time series variance for the
whole sample, so the CLT factors are not directly applicable.
Second, the data from the empirical price process are, in all
likelihood, not drawn from a Brownian semimartingale (e.g., there
are jumps and microstructure frictions). $RRV_{m}^{ \Delta}$, in
turn, behaves differently for other specifications, which we address
elsewhere.

The correlation between $RV^{ \Delta}$ and $RRV_{m}^{ \Delta}$ is
0.982, pointing towards little gain - at relevant frequencies - from
taking linear combinations of the estimators to further reduce
sampling variation. From the joint asymptotic distribution of $(RV^{
\Delta}, RRV_{m}^{ \Delta})$, the conditional covariance matrix at
time $u$ is given by:
\begin{equation}
\Sigma_{u} = \sigma_{u}^{4} \left(
\begin{array}{cc}
2 & \\[0.25cm]
\displaystyle \frac{ \text{cov} \left( W_{1}^{2}, s_{W, m}^{2}
\right)}{ \lambda_{2, m}} \ & \Lambda_{m}
\end{array} \right).
\end{equation}
The covariance term appearing in $\Sigma_{u}$ is hard to tackle
analytically. In unreported results, we used simulations to inspect
the structure of the correlation coefficient around a grid of values
for $m$ that matches our sample. Based on this, we found that the
estimated empirical correlation is slightly higher than the
theoretical level.

\begin{figure}
    \begin{centering}
            \includegraphics[height=9.00cm,width=\linewidth]
                {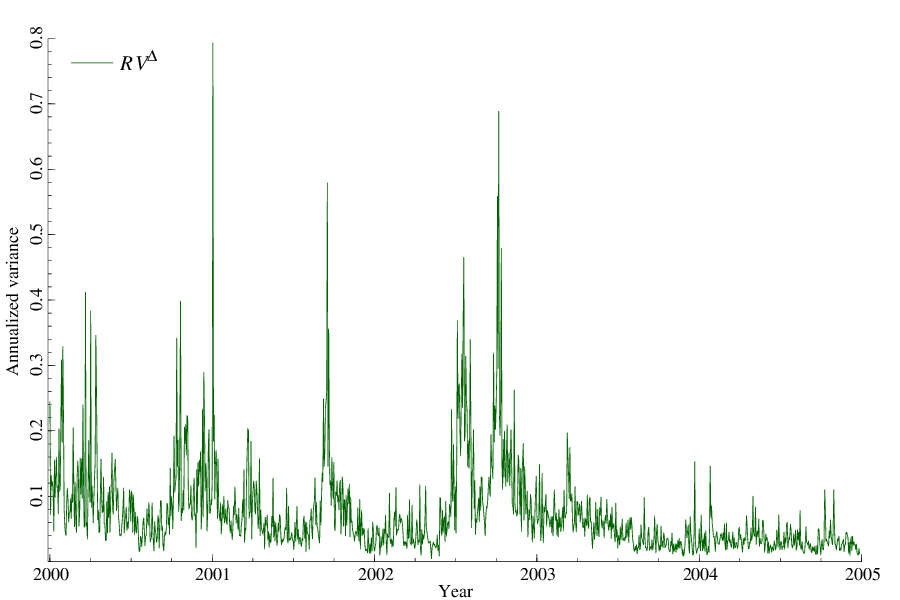}
            \includegraphics[height=9.00cm,width=\linewidth]
                {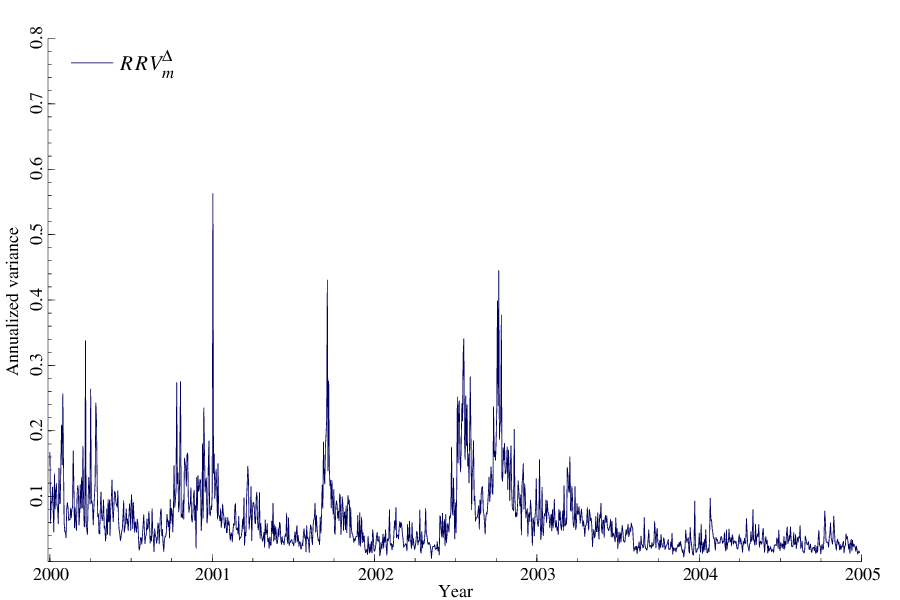}\\
    \end{centering}
    \noindent \caption{The time series of $RV^{ \Delta}$ and
    $RRV_{m}^{ \Delta}$ for GM are shown through the sample period
    January 3, 2000 to December 31, 2004; or 1,255 trading days in
    total. These are constructed from 5-minute midquote returns or
    ranges for data extracted from TAQ.}
    \label{GM-qu-Ts.eps}
\end{figure}

\begin{center}
[ INSERT FIGURE \ref{GM-qu-Ts.eps} ABOUT HERE ]
\end{center}

In Figure \ref{GM-qu-Ts.eps}, $IV$ estimates are drawn for the two
methods, $RV^{ \Delta}$ and $RRV_{m}^{ \Delta}$. The time series
agree on the level of $IV$. The key point is that the sample path of
$RRV_{m}^{ \Delta}$ is less volatile compared to $RV^{ \Delta}$ (but
still appears quite erratic). Again, this suggests that the sampling
errors of $RV^{ \Delta}$ are larger compared to $RRV_{m}^{ \Delta}$,
and that the theoretical gains of the realized range-based estimator
also hold for the empirical identification of $\sigma$, at least for
the 5-minute frequency.

\begin{figure}
    \begin{centering}
            \includegraphics[height=9.00cm,width=\linewidth]
                {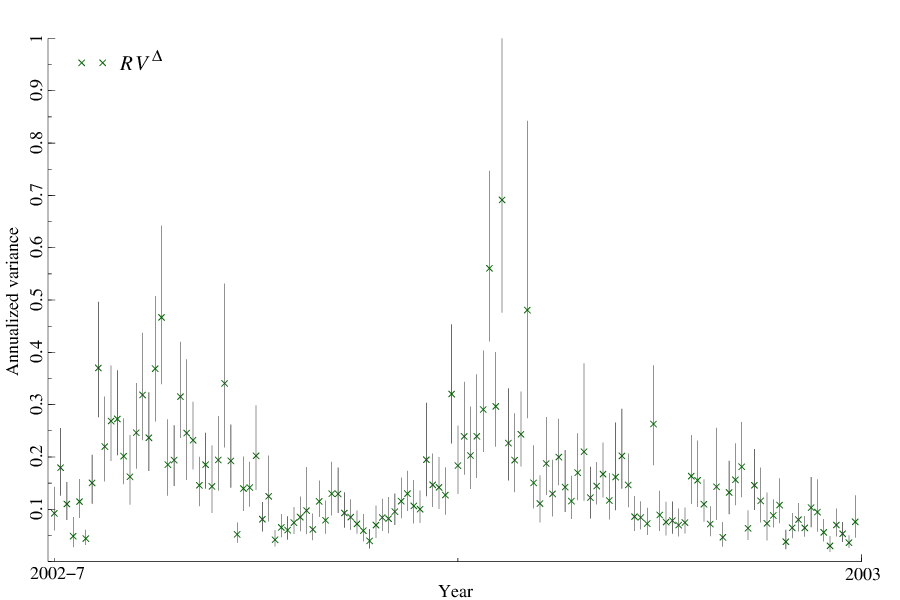}
            \includegraphics[height=9.00cm,width=\linewidth]
                {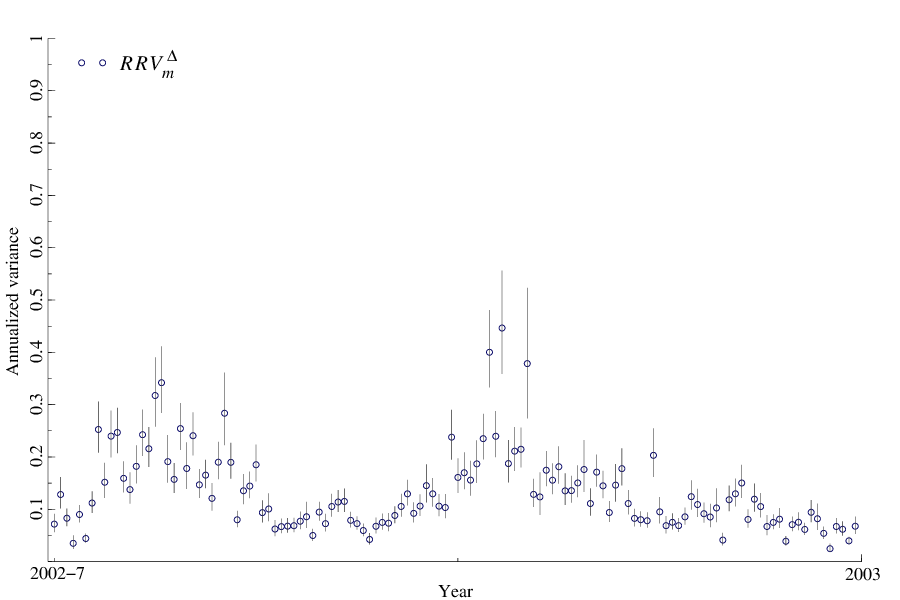}\\
    \end{centering}
    \noindent \caption{$RV^{ \Delta}$ and $RRV_{m}^{ \Delta}$ are
    drawn with 95\% confidence intervals (vertical lines) for the
    period July 1, 2002 through December 31, 2002. The bands
    are constructed from the log-based limit theory in Equation
    \eqref{RbPCiDLn}.}
    \label{GM-qu-TswCI.eps}
\end{figure}

\begin{center}
[ INSERT FIGURE \ref{GM-qu-TswCI.eps} ABOUT HERE ]
\end{center}

To underscore these insights, we extracted data from July 1, 2002 to
December 31, 2002 to plot the $IV$ estimates in Figure
\ref{GM-qu-TswCI.eps} together with 95\% confidence intervals,
constructed from the log-based theory. The confidence bands widen as
expected, when $\sigma$ goes up. Nonetheless, the stability of
$RRV_{m}^{ \Delta}$ feeds into much smaller intervals, consistent
with the theoretical relationship between the $m$ and $\Lambda_{m}$
scalars from Figure \ref{CLT.eps}. This implies that very few
increments are required for $RRV_{m}^{ \Delta}$ to gain a
significant advantage in efficiency over $RV^{
\Delta}$.\footnote{With $\# \Delta r_{ \tau_{i}} \neq 0$ equal to
1,017 on average for the midquote data, we have roughly $m = 13$
increments within each of the seventy eight 5-minute intervals
during the trading day.}

\begin{figure}
    \begin{centering}
            \includegraphics[height=10.00cm,width=\linewidth]
                {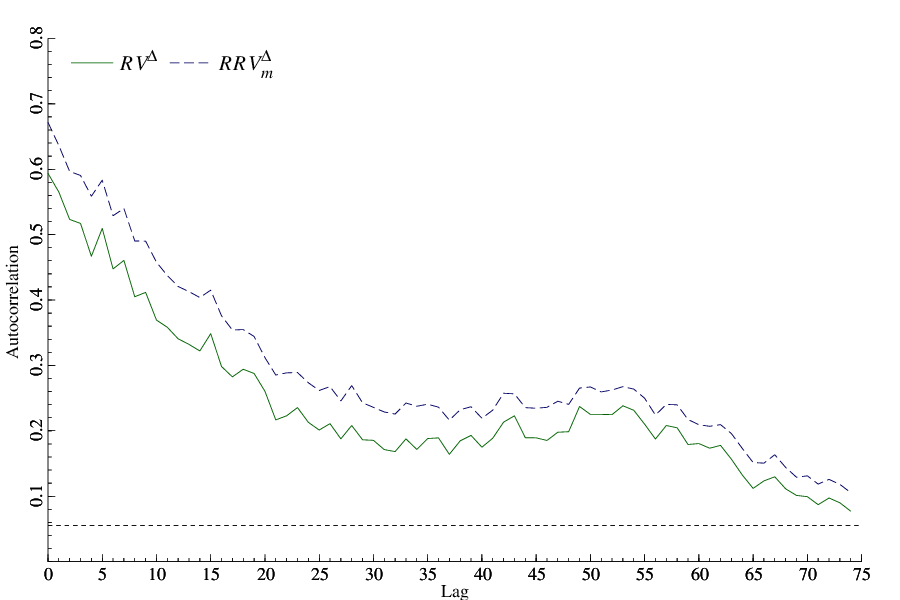}\\
    \end{centering}
    \noindent \caption{Autocorrelation functions of $RV^{ \Delta}$
    and $RRV_{m}^{ \Delta}$. The first 75 lags are shown and the
    dashed horizontal line is Bartlett's standard error for testing a
    white noise hypothesis.}
    \label{GM-qu-Acf.eps}
\end{figure}

\begin{center}
[ INSERT FIGURE \ref{GM-qu-Acf.eps} ABOUT HERE ]
\end{center}

These empirical findings translate into a more persistent time
series behavior for $RRV_{m}^{ \Delta}$, as shown by the
autocorrelation functions in Figure \ref{GM-qu-Acf.eps}. We included
the first 75 lags and report Bartlett two standard error bands for
testing a white noise null hypothesis. All autocorrelations are
positive, starting at about 0.60 - 0.70 and ending around 0.10 -
0.15. The decay pattern in the series is identical but it evolves
more smoothly and at higher levels for $RRV_{m}^{ \Delta}$.
Combined, these observations might be put to work in a forecasting
exercise, although we do not pursue this idea here.

All told, realized range-based estimation of $IV$ offers several
advantages compared to $RV$, both from a theoretical and practical
viewpoint. We acknowledge, however, that the probabilistic theory
proposed in this paper needs further refinement at higher
frequencies, where microstructure noise is more problematic.
Statistical tools for controlling the impact of such noise is
crucial for getting consistent estimates of $IV$. These techniques
have already been developed for $RV$, see, e.g.,
\cite{zhang-mykland-ait-sahalia:05a} or
\cite{barndorff-nielsen-hansen-lunde-shephard:08a}. It presents a
topic for future research to verify if our method extends along
these lines, and we are currently undertaking a formal analysis of
$RRV$ and market microstructure noise.

\section{Conclusions and directions for future research}
\textit{Realized range-based variance} ($RRV$) is an approach based
on intraday price ranges for non-parametric measurement of the
\textit{integrated variance} ($IV$) of continuous semimartingales.
Under weak regularity conditions, we have shown that it can extract
$IV$ more accurately than previous methods, when microstructure
noise is preventing \textit{realized variance} ($RV$) from being
sampled at the maximum frequency. Another contribution of this
paper, particularly useful in empirical analysis, is the solution to
the downward bias problem that has haunted the range-based
literature for decades.

The finite sample distributions of the estimator were inspected with
Monte Carlo analysis. For moderate samples, the coverage
probabilities of the confidence bands for the t-statistics
correspond with the limit theory, in particular for log-based
inference.

We highlighted the empirical potential of $RRV$ vis-\`{a}-vis $RV$
by applying our method to a set of high-frequency data for General
Motors. Consistent with the theory, $RRV$ has smaller confidence
bands than $RV$. Although empirical price processes are very
different from diffusion models and real data are noisy objects, we
feel the results support our theory quite well and opens up
alternative routes for estimating $IV$.

In future projects, we envision several extensions of the current
framework. First, there is plenty of evidence against the continuous
sample path diffusion. We are convinced that a range-based statistic
can estimate \textit{quadratic variation} ($QV$), when the price
also exhibits jumps. This theory is being developed in
\cite{christensen-podolskij:12a}, along with realized range-based
bi-power variation. Second, with microstructure noise in observed
asset prices, further comparisons of $RRV$ and $RV$ are needed.
Finally, we can handle the bivariate case with the polarization
identities, so multivariate range-based analysis constitutes a
promising future application.
    \pagebreak

\appendix
\section{Appendix}
Without loss of generality, in the following we restrict the
functions $\mu$ and $\sigma$ to be bounded (e.g.,
\cite{barndorff-nielsen-graversen-jacod-podolskij-shephard:06a}).

\subsection{Proof of Theorem \ref{RbPCiP}}
First, define:
\begin{align*}
\xi_{i}^{n} &= \frac{1}{ \lambda_{2}} \sigma^{2}_{ \frac{i - 1}{n}}
s_{W_{i \Delta, \Delta}}^{2}, \\[0.25cm]
U_{n} &= \sum_{i = 1}^{n} \xi_{i}^{n},
\end{align*}
and note that:
\begin{equation*}
\mathbb{E} \left[ \xi_{i}^{n} \mid \mathcal{F}_{ \frac{i - 1}{n}}
\right] = \frac{1}{n} \sigma^{2}_{ \frac{i - 1}{n}}.
\end{equation*}
So
\begin{equation}
\label{ApI} \sum_{i = 1}^{n} \mathbb{E} \left[ \xi_{i}^{n} \mid
\mathcal{F}_{ \frac{i - 1}{n}} \right] \overset{p}{ \to} IV.
\end{equation}
Now, by setting
\begin{equation*}
\eta_{i}^{n} = \xi_{i}^{n} - \mathbb{E} \left[ \xi_{i}^{n} \mid
\mathcal{F}_{ \frac{i - 1}{n}} \right],
\end{equation*}
we get:
\begin{equation*}
\mathbb{E} \left[ \left( \eta_{i}^{n} \right)^{2} \mid \mathcal{F}_{
\frac{i - 1}{n}} \right] = \Lambda \frac{1}{n^{2}} \sigma^{4}_{
\frac{i - 1}{n}}.
\end{equation*}
Therefore,
\begin{equation*}
\sum_{i = 1}^{n} \mathbb{E} \left[ \left( \eta_{i}^{n} \right)^{2}
\mid \mathcal{F}_{ \frac{i - 1}{n}} \right] \overset{p}{ \to} 0.
\end{equation*}
Hence, $U_{n} \overset{p}{ \to} IV$ follows from \eqref{ApI}. As a
sufficient condition in the next step, we deduce that $RRV^{ \Delta}
- U_{n} \overset{p}{ \to} 0$. Note the equality
\begin{align*}
\label{DifLPW} RRV^{ \Delta} - U_{n} &= \frac{1}{ \lambda_{2}}
\sum_{i = 1}^{n} \left( s_{p_{i \Delta, \Delta}} - \sigma_{ \frac{i
- 1}{n}} s_{W_{i \Delta, \Delta}} \right) \left( s_{p_{i \Delta,
\Delta}} + \sigma_{ \frac{i - 1}{n}} s_{W_{i \Delta, \Delta}}
\right) \\[0.25cm]
&\equiv R_{n}^{1} + R_{n}^{2},
\end{align*}
with $R_{n}^{1}$ and $R_{n}^{2}$ defined by:
\begin{align*}
R_{n}^{1} &= \frac{2}{ \lambda_{2}} \sum_{i = 1}^{n} \sigma_{
\frac{i - 1}{n}} s_{W_{i \Delta, \Delta}} \left( s_{p_{i \Delta,
\Delta}} - \sigma_{ \frac{i - 1}{n}} s_{W_{i \Delta, \Delta}}
\right), \\[0.25cm]
R_{n}^{2} &= \frac{1}{ \lambda_{2}} \sum_{i = 1}^{n} \left( s_{p_{i
\Delta, \Delta}} - \sigma_{ \frac{i - 1}{n}} s_{W_{i \Delta,
\Delta}} \right)^{2}.
\end{align*}
We decompose the second term further:
\begin{align*}
R_{n}^{2} & \leq \frac{1}{ \lambda_{2}} \sum_{i = 1}^{n} \Biggl(
\underset{ \left(i - 1 \right) / n \leq s, t \leq i / n}{ \sup \mid
\int_{s}^{t}} \mu_{u} \text{d}u + \int_{s}^{t} \left( \sigma_{u} -
\sigma_{ \frac{i - 1}{n}} \right) \text{d}W_{u} \mid \Biggr)^{2}
\\[0.25cm]
&\leq \frac{2}{ \lambda_{2}} \sum_{i = 1}^{n} \Biggl( \underset{
\left(i - 1 \right)/n \leq s, t \leq i/n}{ \sup \mid \int_{s}^{t}}
\mu_{u} \text{d}u \mid \Biggr)^{2} + \frac{2}{\lambda_{2}} \sum_{i =
1}^{n} \Biggl( \underset{ \left(i - 1 \right)/n \leq s, t \leq i/n}{
\sup \mid \int_{s}^{t}} \left( \sigma_{u} - \sigma_{ \frac{i -
1}{n}} \right) \text{d}W_{u} \mid \Biggr)^{2} \\[0.25cm]
&\equiv R_{n}^{2.1} + R_{n}^{2.2}.
\end{align*}
It is straightforward to verify that $\mathbb{E} \left[ R_{n}^{2.1}
 \right] = \text{O} \left( n^{-1} \right)$. For the latter term, we
exploit the Burkholder inequality (e.g., \cite{revuz-yor:98a}):
\begin{align*}
\mathbb{E} \left[ R_{n}^{2.2} \right] & \leq \frac{2C}{ \lambda_{2}}
\sum_{i = 1}^{n} \mathbb{E} \left[ \int_{ \frac{i - 1}{n}}^{
\frac{i}{n}} \left( \sigma_{u} - \sigma_{ \frac{i - 1}{n}}
\right)^{2} \text{d}u \right] \\[0.25cm]
&= \frac{2C}{ \lambda_{2}} \mathbb{E} \left[ \int_{0}^{1} \left(
\sigma_{u} - \sigma_{ \frac{ \left[ nu \right]}{n}} \right)^{2}
\text{d}u \right] \\[0.25cm]
&= \text{o} \left( 1 \right),
\end{align*}
for some constant $C > 0$. Thus, $R_{n}^{2} = \text{o}_{p} \left( 1
\right)$. Using a decomposition as above and the Cauchy-Schwarz
inequality, we have that $R_{n}^{1} = \text{o}_{p} \left( 1
\right)$. By collecting terms, $RRV^{ \Delta} - U_{n} \overset{p}{
\to} 0$. \hfill {$\blacksquare$}
\subsection{Proof of Theorem \ref{RbPCiD}}
We need the following lemma.
\begin{lemma}
\label{Unique} Given two continuous functions $f, g : I \to
\mathbb{R}$ on compact $I \subseteq \mathbb{R}^{n}$, assume $t^{*}$
is the only point in $I$ where the maximum of $f$ is achieved. Then
it holds:
\begin{equation*}
M_{\epsilon} \left( g \right) \equiv \frac{1}{\epsilon} \left[
\sup_{t \in I} \left\{ f \left( t \right) + \epsilon g \left( t
\right) \right\} - \sup_{t \in I} \left\{ f \left( t \right)
\right\} \right] \to g \left( t^{*} \right) \text{\quad as \quad}
\epsilon \downarrow 0.
\end{equation*}
\end{lemma}
\textbf{Proof} \\[0.25cm]
Construct the set
\begin{equation*}
\bar{G} = \Bigl\{ h \in C \left( I \right) \mid h \text{\textup{ is
constant on }} B_{ \delta} \left( t^{*} \right) \cap I
\text{\textup{ for some }} \delta > 0 \Bigr\}.
\end{equation*}
As usual, $C \left( I \right)$ is the set of continuous functions on
$I$ and $B_{ \delta} \left( t^{*} \right)$ is an open ball of radius
$\delta$ centered at $t^{*}$. Take $\bar{g} \in \bar{G}$ and recall
$\bar{g}$ is bounded on $I$. Thus, for $\epsilon$ sufficiently
small:
\begin{align*}
\sup_{t \in I} \left\{ f \left( t \right) + \epsilon \bar{g} \left(
t \right) \right\} &= \max \left\{ \underset{t \in I \cap B_{
\delta} \left( t^{*} \right)}{ \sup \left\{ f \left( t \right) +
\epsilon \bar{g} \left( t \right) \right\}}, \underset{t \in I \cap
B_{ \delta}^{c} \left( t^{*} \right)}{ \sup \left\{ f \left( t
\right) + \epsilon \bar{g} \left( t \right) \right\}} \right\}
\\[0.25cm]
&= \underset{t \in I \cap B_{ \delta} \left( t^{*} \right)}{ \sup
\left\{ f \left( t \right) + \epsilon \bar{g} \left( t \right)
\right\}} \\[0.25cm]
&= f \left( t^{*} \right) + \epsilon \bar{g} \left( t^{*} \right).
\end{align*}
So,
\begin{equation*}
M_{ \epsilon} \left( \bar{g} \right) \to \bar{g} \left( t^{*}
\right),
\end{equation*}
$\forall \ \bar{g} \in \bar{G}$. Now, let $g \in C \left( I
\right)$. As $\bar{G}$ is dense in $C \left( I \right)$, $\exists \
\bar{g} \in \bar{G} : \bar{g} \left( t^{*} \right) = g \left( t^{*}
\right)$ and $\left| \bar{g} - g \right|_{ \infty} < \epsilon^{
\prime}$ ($ \left| \, \cdot \, \right|_{ \infty}$ is the sup-norm).
We see that $\left| M_{ \epsilon} \left( \bar{g} \right) - M_{
\epsilon} \left( g \right) \right| < \epsilon^{ \prime}$, and
\begin{equation*}
\left| M_{ \epsilon} \left( g \right) - g \left( t^{*} \right)
\right| \leq \left| M_{ \epsilon} \left( \bar{g} \right) - \bar{g}
\left( t^{*} \right) \right| + \left| M_{ \epsilon} \left( g \right)
- M_{ \epsilon} \left( \bar{g} \right) \right| \to 0.
\end{equation*}
Thus, the assertion is established.\hfill $\blacksquare$\\[-0.25cm]

\noindent With this lemma at hand, we proceed with a three-stage
proof of Theorem \ref{RbPCiD}. In the first part, a CLT is proved
for the quantity
\begin{equation*}
\bar{U}_{n} = \sqrt{n} \sum_{i = 1}^{n} \eta_{i}^{n}.
\end{equation*}
The second step is to define a new sequence:
\begin{equation*}
U_{n}^{ \prime} = \sqrt{n} \frac{1}{ \lambda_{2}} \sum_{i = 1}^{n}
\left( s_{p_{i \Delta, \Delta}}^{2} - \mathbb{E} \left[ s_{p_{i
\Delta, \Delta}}^{2} \mid \mathcal{F}_{ \frac{i - 1}{n}} \right]
\right),
\end{equation*}
and show the result
\begin{equation*}
U_{n}^{ \prime} - \bar{U}_{n} \overset{p}{ \to} 0.
\end{equation*}
The interested reader may note that assumption \textbf{(V)} is not
needed for Part I and II. Finally, in Part III, the theorem follows
from:
\begin{align*}
& \sqrt{n} \sum_{i = 1}^{n} \left( \frac{1}{\lambda_{2}} \mathbb{E}
\left[ s_{p_{i \Delta, \Delta}}^{2} \mid \mathcal{F}_{ \frac{i -
1}{n}} \right] - \mathbb{E} \left[ \xi_{i}^{n} \mid \mathcal{F}_{
\frac{i - 1}{n}} \right] \right) \overset{p}{ \to} 0, \text{ and}
\\[0.25cm]
& \sqrt{n} \left( \sum_{i = 1}^{n} \mathbb{E} \left[ \xi_{i}^{n}
\mid \mathcal{F}_{ \frac{i - 1}{n}} \right] - IV \right)
\overset{p}{ \to} 0.
\end{align*}
\textbf{Proof of part I} \\[0.25cm]
Notice that:
\begin{equation*}
n \sum_{i = 1}^{n} \mathbb{E} \left[ \left( \eta_{i}^{n} \right)^{2}
\mid \mathcal{F}_{ \frac{i - 1}{n}} \right] \overset{p}{ \to}
\Lambda IQ,
\end{equation*}
and by the scaling property of Brownian motion,
\begin{equation*}
\sqrt{n} \sum_{i = 1}^{n} \mathbb{E} \left[ \eta_{i}^{n} \left( W_{
\frac{i}{n}} - W_{ \frac{i - 1}{n}} \right) \mid \mathcal{F}_{
\frac{i - 1}{n}} \right] \overset{p}{ \to} \frac{ \nu}{ \lambda_{2}}
IV,
\end{equation*}
where $\nu = \mathbb{E} \left[ W_{1} s_{W}^2 \right]$. As $W
\overset{d}{=} - W$, it follows that $\nu = - \nu $ and, hence, $\nu
= 0$.

Next, let $N = \left( N_{t} \right)_{t \in \left[ 0, 1 \right]}$ be
a bounded martingale on $\bigl( \Omega, \mathcal{F}, \left(
\mathcal{F}_{t} \right)_{t \in \left[ 0, 1 \right]}, P \bigr)$,
which is orthogonal to $W$ (i.e., with quadratic covariation
$\left\langle \, W, N \, \right\rangle_{t} = 0$, almost surely).
Then:
\begin{equation}
\label{E(Inc_N)} \sqrt{n} \sum_{i = 1}^{n} \mathbb{E} \left[
\eta_{i}^{n} \left( N_{ \frac{i}{n}} - N_{ \frac{i - 1}{n}} \right)
\mid \mathcal{F}_{ \frac{i - 1}{n}} \right] = 0.
\end{equation}
For this result, we use Clark's Representation Theorem (see, e.g.,
\cite[Appendix E]{karatzas-shreve:98a}):
\begin{equation}
\label{ClarkRT} s_{W_{i \Delta, \Delta}}^{2} - \frac{1}{n}
\lambda_{2} = \int_{ \frac{i - 1}{n}}^{ \frac{i}{n}} H_{u}^{n}
\text{d}W_{u},
\end{equation}
for some predictable function $H_{u}^{n}$. Notice $\mathbb{E} \left[
\int_{a}^{b} f_{u} \text{d}W_{u} \left( N_{b} - N_{a} \right) \mid
\mathcal{F}_{a} \right] = 0$, for any $\left[ a, b \right]$ and
predictable $f$. To prove this assertion, take a partition $a =
t_{0}^{*} < t_{1}^{*} < \ldots < t_{n}^{*} = b$ and compute:
\begin{align*}
& \mathbb{E} \left[ \sum_{i = 1}^{n} f_{t_{i - 1}^{*}} \left(
W_{t_{i}^{*}} - W_{t_{i - 1}^{*}} \right) \left( N_{b} - N_{a}
\right) \mid \mathcal{F}_{a} \right] = \mathbb{E} \left[ \sum_{i =
1}^{n} f_{t_{i - 1}^{*}} \left( W_{t_{i}^{*}} - W_{t_{i - 1}^{*}}
\right) N_{b} \mid \mathcal{F}_{a} \right] \\[0.25cm]
& = \mathbb{E} \left[ \sum_{i = 1}^{n} \mathbb{E} \left[ \mathbb{E}
\left[ f_{t_{i - 1}^{*}} \left( W_{t_{i}^{*}} - W_{t_{i - 1}^{*}}
\right) N_{b} \mid \mathcal{F}_{t_{i}^{*}} \right] \mid
\mathcal{F}_{t_{i - 1}^{*}} \right] \mid \mathcal{F}_{a} \right]
\\[0.25cm]
& = 0.
\end{align*}
From Equation \eqref{ClarkRT}, \eqref{E(Inc_N)} is attained.
Finally, stable convergence in law follows by Theorem IX 7.28 in
\cite{jacod-shiryaev:03a}:
\begin{equation*}
\bar{U}_{n} \overset{d_{s}}{ \to} \sqrt{ \Lambda} \int_{0}^{1}
\sigma_{u}^{2} \text{d}B_{u}.
\end{equation*}
\hfill $\blacksquare$ \\
\textbf{Proof of part II} \\[0.25cm]
We begin by setting
\begin{equation*}
\zeta_{i}^{n} = \sqrt{n} \left( \frac{1}{ \lambda_{2}} s_{p_{i
\Delta, \Delta}}^{2} - \xi_{i}^{n} \right),
\end{equation*}
and obtain the identity:
\begin{equation*}
U_{n}^{ \prime} - \bar{U}_{n} = \sum_{i = 1}^{n} \left(
\zeta_{i}^{n} - \mathbb{E} \left[ \left( \zeta_{i}^{n} \right)^{2}
\mid \mathcal{F}_{ \frac{i - 1}{n}} \right] \right).
\end{equation*}
To complete the second step, it suffices that
\begin{equation*}
\sum_{i = 1}^{n} \mathbb{E} \left[ \left( \zeta_{i}^{n} \right)^{2}
\right] \to 0.
\end{equation*}
We can show this result with the same methods applied to the
estimates of $R_{n}^{1}$ and $R_{n}^{2}$ in the proof of Theorem
\ref{RbPCiP}. \hfill $\blacksquare$ \\[0.50cm]
\textbf{Proof of part III} \\[0.25cm]
It holds that:
\begin{equation*}
\sqrt{n} \left( \sum_{i = 1}^{n} \mathbb{E} \left[ \xi_{i}^{n} \mid
\mathcal{F}_{ \frac{i - 1}{n}} \right] - IV \right) = \sqrt{n}
\sum_{i = 1}^{n} \int_{ \frac{i - 1}{n}}^{ \frac{i}{n}} \left(
\sigma_{ \frac{i - 1}{n}}^{2} - \sigma_{u}^{2} \right) \text{d}u.
\end{equation*}
Exploiting the results of
\cite{barndorff-nielsen-graversen-jacod-podolskij-shephard:06a}, we
find that, under assumption V$_{2}$,
\begin{equation*}
\sqrt{n} \left( \sum_{i = 1}^{n} \mathbb{E} \left[ \xi_{i}^{n} \mid
\mathcal{F}_{ \frac{i - 1}{n}} \right] - IV \right) \overset{p}{
\to} 0.
\end{equation*}
Now, we prove the first convergence of Part III stated above. After
some computations - identical to the methods in Theorem \ref{RbPCiP}
- we get, using V$_{2}$,
\begin{align*}
& \sqrt{n} \sum_{i = 1}^{n} \left( \frac{1}{ \lambda_{2}} \mathbb{E}
\left[ s_{p_{i \Delta, \Delta}}^{2} \mid \mathcal{F}_{ \frac{i -
1}{n}} \right] - \mathbb{E} \left[ \xi_{i}^{n} \mid \mathcal{F}_{
\frac{i - 1}{n}} \right] \right) \\[0.25cm]
& = \sqrt{n} \frac{2}{ \lambda_{2}} \sum_{i = 1}^{n} \mathbb{E}
\left[ \sigma_{ \frac{i - 1}{n}} s_{W_{i \Delta, \Delta}} \left(
s_{p_{i \Delta, \Delta} - \sigma_{ \frac{i - 1}{n}}} s_{W_{ i
\Delta, \Delta}} \right) \mid \mathcal{F}_{ \frac{i - 1}{n}} \right]
+ \text{o}_{p} \left( 1 \right) \\[0.25cm]
& = \sqrt{n} \frac{2}{ \lambda_{2}} \sum_{i = 1}^{n} \mathbb{E}
\Biggl[ \sigma_{ \frac{i - 1}{n}} s_{W_{i \Delta, \Delta}} \Biggl(
\underset{ \left( i - 1 \right) / n \leq s, t \leq i / n}{ \sup
\biggl( \sigma_{ \frac{i - 1}{n}} ( W_{t} } - W_{s} ) + \int_{s}^{t}
\mu_{u} \text{d}u + \int_{s}^{t} \left( \sigma_{u} - \sigma_{
\frac{i - 1}{n}} \right) \text{d}W_{u} \biggr) \\[0.25cm]
& - \sigma_{ \frac{i - 1}{n}} s_{W_{i \Delta, \Delta}} \Biggr) \mid
\mathcal{F}_{ \frac{i - 1}{n}} \Biggr] + \text{o}_{p} \left( 1
\right).
\end{align*}
By appealing to assumption V$_{2}$ again, we get the decomposition:
\begin{equation*}
\sqrt{n} \sum_{i = 1}^{n} \left( \frac{1}{ \lambda_{2}} \mathbb{E}
\left[ s_{p_{i \Delta, \Delta}}^{2} \mid \mathcal{F}_{ \frac{i -
1}{n}} \right] - \mathbb{E} \left[ \xi_{i}^{n} \mid \mathcal{F}_{
\frac{i - 1}{n}} \right] \right) = V_{n}^{1} + V_{n}^{2} +
\text{o}_{p} \left( 1 \right),
\end{equation*}
with the random variables $V_{n}^{1}$ and $V_{n}^{2}$ defined by
\begin{align*}
V_{n}^{1} &= \frac{2}{ \lambda_{2}} \sum_{i = 1}^{n} \mathbb{E}
\biggl[ \sigma_{ \frac{i - 1}{n}} s_{W_{i \Delta, \Delta}} \biggl\{
\underset{ \left( i - 1 \right) / n \leq s, t \leq i / n}{ \sup
\biggl( \sqrt{n} \sigma_{ \frac{i - 1}{n}}} \left( W_{t} - W_{s}
\right) + \sqrt{n} \int_{s}^{t} \mu_{ \frac{i - 1}{n}}
\text{d}u \\[0.25cm]
&+ \sqrt{n} \int_{s}^{t} \left\{ \sigma_{ \frac{i - 1}{n}}^{ \prime}
\left( W_{u} - W_{ \frac{i - 1}{n}} \right) + v_{ \frac{i - 1}{n}}
\left( B_{u}^{ \prime} - B_{ \frac{i - 1}{n}}^{ \prime} \right)
\right\} \text{d}W_{u} - \sqrt{n} \sigma_{ \frac{i - 1}{n}} s_{W_{i
\Delta, \Delta}} \biggr\} \mid \mathcal{F}_{ \frac{i - 1}{n}}
\biggr],
\end{align*}
and
\begin{align*}
V_{n}^{2} &= \sqrt{n} \frac{2}{ \lambda_{2}} \sum_{i = 1}^{n}
\mathbb{E} \biggl[ \sigma_{ \frac{i - 1}{n}} s_{W_{i \Delta,
\Delta}} \biggl\{ \underset{ \left( i - 1 \right) / n \leq s, t \leq
i / n}{ \sup \biggl( \sigma_{ \frac{i - 1}{n}} ( W_{t}} - W_{s} ) +
\int_{s}^{t} \mu_{u} \text{d}u + \int_{s}^{t} \left( \sigma_{u} -
\sigma_{ \frac{i - 1}{n}} \right) \text{d}W_{u} \biggr) \\[0.25cm]
&- \sigma_{ \frac{i - 1}{n}} s_{W_{i \Delta, \Delta}} \biggr\} \mid
\mathcal{F}_{ \frac{i - 1}{n}} \biggr] - V_{n}^{1} \\[0.25cm]
& \leq \sqrt{n} \frac{2}{ \lambda_{2}} \sum_{i = 1}^{n} \mathbb{E}
\biggl[ \sigma_{ \frac{i - 1}{n}} s_{W_{i \Delta, \Delta}} \biggl\{
\underset{ \left( i - 1 \right) / n \leq s, t \leq i / n}{ \sup
\biggl( \int_{s}^{t} \Bigl( \mu_{u} - } \mu_{ \frac{i - 1}{n}}
\Bigr) \text{d}u + \int_{s}^{t} \biggl\{ \int_{ \frac{i - 1}{n}}^{u}
\mu_{r}^{ \prime} \text{d}r \\[0.25cm]
&+ \int_{ \frac{i - 1}{n}}^{u} \left( \sigma_{r}^{ \prime} -
\sigma_{ \frac{i - 1}{n}}^{ \prime} \right) \text{d}W_{r} + \int_{
\frac{i - 1}{n}}^{u} \left( v_{r}^{ \prime} - v_{ \frac{i - 1}{n}}^{
\prime} \right) \text{d}B_{r}^{ \prime} \biggr\} \text{d}W_{u}
\biggr) \biggr\} \mid \mathcal{F}_{ \frac{i - 1}{n}} \biggr].
\end{align*}
From the Cauchy-Schwarz and Burkholder inequalities, we find that
\begin{equation*}
V_{n}^{2} = \text{o}_{p} \left( 1 \right).
\end{equation*}
At this point, we invoke Lemma \ref{Unique} by setting:
\begin{align*}
f_{in} \left( s, t \right) &= \sqrt{n} \sigma_{ \frac{i - 1}{n}}
\left( W_{t} - W_{s} \right), \\[0.25cm]
g_{in} \left( s, t \right) &= n \int_{s}^{t} \mu_{ \frac{i - 1}{n}}
\text{d}u + n \int_{s}^{t} \left\{ \sigma_{ \frac{i - 1}{n}}^{
\prime} \left( W_{u} - W_{ \frac{i - 1}{n}} \right) + v_{ \frac{i -
1}{n}}^{ \prime} \left( B_{u}^{ \prime} - B_{ \frac{i - 1}{n}}^{
\prime} \right) \right\} \text{d}W_{u}
\\[0.25cm]
&= \mu_{ \frac{i - 1}{n}} g_{in}^{1} \left( s, t \right) + \sigma_{
\frac{i - 1}{n}}^{ \prime} g_{in}^{2} \left( s, t \right) + v_{
\frac{i - 1}{n}}^{ \prime} g_{in}^{3} \left( s, t \right).
\end{align*}
Note that $\epsilon = 1/ \sqrt{n}$. Through assumption V$_{1}$, we
get the following identity:
\begin{align*}
\left( t_{in}^{*} \left( W \right), s_{in}^{*} \left( W \right)
\right) &= \underset{ \left( i - 1 \right) / n \leq s, t \leq i /
n}{ \arg \sup f_{in} \left( s, t \right) } \\[0.25cm]
&= \underset{ \left( i - 1 \right) / n \leq s, t \leq i /
n}{ \arg \sup \sqrt{n} ( W_{t} } - W_{s} ) \\[0.25cm]
&\overset{d}{=} \underset{ 0 \leq s, t \leq 1}{ \arg \sup} \left(
W_{t} - W_{s} \right).
\end{align*}
A standard result then states that the points $t_{in}^{*} \left( W
\right)$ and $s_{in}^{*} \left( W \right)$ are unique, almost
surely, so the lemma applies. Hence, by repeating the proof of the
lemma, we get the decomposition:
\begin{equation*}
V_{n}^{1} = \frac{2}{ \lambda_{2}} \sum_{i = 1}^{n} \mathbb{E}
\left[ \sigma_{ \frac{i - 1}{n}} s_{W_{i \Delta, \Delta}} \left(
\frac{1}{ \sqrt{n}} g_{in} \left( t^{*}_{in} \left( W \right),
s^{*}_{in} \left( W \right) \right) + R_{in} \right) \mid
\mathcal{F}_{ \frac{i - 1}{n}} \right],
\end{equation*}
where the term $R_{in}$ satisfies:
\begin{equation*}
\mathbb{E} \left[ \left( R_{in} \right)^{2} \right] = \text{o}
\left( n^{-1} \right),
\end{equation*}
(uniformly in $i$). By the Cauchy-Schwarz inequality, we have the
estimation:
\begin{equation*}
\frac{2}{ \lambda_{2}} \sum_{i = 1}^{n} \mathbb{E} \left[ \sigma_{
\frac{i - 1}{n}} s_{W_{i \Delta, \Delta}} R_{in} \mid \mathcal{F}_{
\frac{i - 1}{n}} \right] = \text{o}_{p} \left( 1 \right).
\end{equation*}
As $g_{in}^{1} \left( s, t \right)$, $g_{in}^{2} \left( s, t
\right)$ and $g_{in}^{3} \left( s, t \right)$ are independent of
$\mathcal{F}_{ \frac{i - 1}{n}}$, we obtain:
\begin{align*}
\mathbb{E} \left[ \sigma_{ \frac{i - 1}{n}} s_{W_{i \Delta, \Delta}}
\frac{1}{ \sqrt{n}} g_{in}^{1} \left( t_{in}^{*} \left( W \right),
s_{in}^{*} \left( W \right) \right) \mid \mathcal{F}_{ \frac{i -
1}{n}} \right] & \equiv \frac{1}{ \sqrt{n}} \sigma_{ \frac{i -
1}{n}} \mu_{ \frac{i - 1}{n}}
\nu_{1} \\[0.25cm]
\mathbb{E} \left[ \sigma_{ \frac{i - 1}{n}} s_{W_{i \Delta, \Delta}}
\frac{1}{ \sqrt{n}} g_{in}^{2} \left( t_{in}^{*} \left( W \right),
s_{in}^{*} \left( W \right) \right) \mid \mathcal{F}_{ \frac{i -
1}{n}} \right] & \equiv \frac{1}{ \sqrt{n}} \sigma_{ \frac{i -
1}{n}} \sigma_{ \frac{i - 1}{n}}^{ \prime} \nu_{2} \\[0.25cm]
\mathbb{E} \left[ \sigma_{ \frac{i - 1}{n}} s_{W_{i \Delta, \Delta}}
\frac{1}{ \sqrt{n}} g_{in}^{3} \left( t_{in}^{*} \left( W \right),
s_{in}^{*} \left( W \right) \right) \mid \mathcal{F}_{ \frac{i -
1}{n}} \right] & \equiv \frac{1}{ \sqrt{n}} \sigma_{ \frac{i -
1}{n}} v_{ \frac{i - 1}{n}} \nu_{3},
\end{align*}
with
\begin{equation*}
\nu_{k} = \mathbb{E} \left[ s_{W_{i \Delta, \Delta}} g_{in}^{k}
\left( t_{in}^{*} \left( W \right), s_{in}^{*} \left( W \right)
\right) \right], \quad \text{for } k = 1, 2, \text{ and } 3.
\end{equation*}
Note that,
\begin{equation}
\label{Points} \left( t^{*}_{in} \left( W \right), s^{*}_{in} \left(
W \right) \right) = \left( s^{*}_{in} \left( - W \right), t^{*}_{in}
\left( - W \right) \right).
\end{equation}
Using \eqref{Points} and the relationship $\left(W, B \right)
\overset{d}{=} \left( -W, -B \right)$, it follows that $\nu_{k} = -
\nu_{k}$ and, hence, $\nu_{k} = 0$ for $k = 1$, 2, and 3. This
yields the estimation:
\begin{equation*}
V_{n}^{1} = \text{o}_{p} \left( 1 \right),
\end{equation*}
and the proof is complete.\hfill $\blacksquare$

\subsection{Proof of Theorem \ref{RbPCiDm}}
The result is shown in the same manner as the proofs of Theorem
\ref{RbPCiP} and \ref{RbPCiD}. \hfill $\blacksquare$
    \clearpage

% LIST OF REFERENCES
\nocite{bandi-russell:06a}
\nocite{barndorff-nielsen-shephard:04b}
\nocite{barndorff-nielsen-shephard:05a}
\nocite{barndorff-nielsen-shephard:07a}
\nocite{doornik:02a}

\renewcommand{\baselinestretch}{1.0}
\normalsize
\bibliographystyle{agsm}
\bibliography{UserRef}
    \pagebreak

% TABLES AND FIGURES
\renewcommand{\baselinestretch}{1.3}
\normalsize

\setlength{\tabcolsep}{0.59cm}
\input{tables/GM-DataC.tex}
    \pagebreak

\setlength{\tabcolsep}{0.45cm}
\input{tables/GM-qu-Sample.tex}
    \pagebreak

    \pagebreak

\begin{figure}
    \begin{centering}
        \includegraphics[height=10.00cm,width=\linewidth]
            {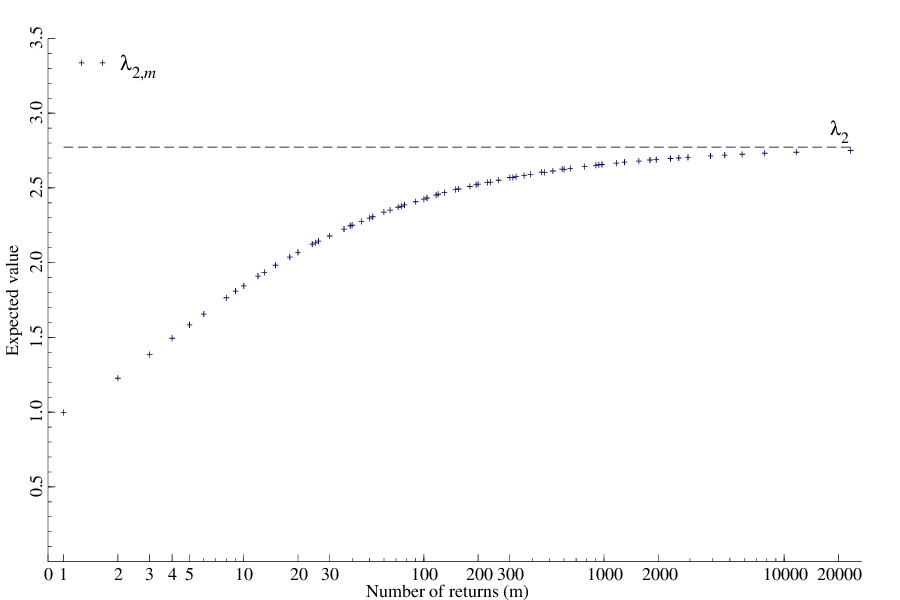}\\
    \end{centering}
    \caption{$\lambda_{2, m}$ against $m$ on a log scale. All
    estimates are from a simulation with 1,000,000 repetitions and
    the dashed line represents the asymptotic value.}
    \label{L(2,m).eps}
\end{figure}
    \pagebreak

    \pagebreak

    \pagebreak

    \pagebreak

    \pagebreak

    \pagebreak
\end{document}

%% file: tables/GM-DataC.tex
\vspace*{\fill}
\begin{table}[H]
\caption{Number of tick data pr. trading day.}
\label{GM-DataC.tex}
\begin{tabular*}{\linewidth}{@{~}*{1}{l}*{7}{r}@{~}}
\\[-0.20cm]
\hline
\rule[-3mm]{0mm}{8mm}Ticker & \multicolumn{3}{c}{Trades} &  & \multicolumn{3}{c}{Quotes}\\
\rule[-3mm]{0mm}{8mm} & All & \#$r_{ \tau_{i}} \neq 0$ & \#$\Delta r_{ \tau_{i}} \neq 0$
 &  & All & \#$r_{ \tau_{i}} \neq 0$ & \#$\Delta r_{ \tau_{i}} \neq 0$\\
GM & 2220 & 960 & 558 &  & 5144 & 1357 & 1017\\
\\[-0.35cm]
\hline
\end{tabular*}
\vspace*{0.15cm}

The table contains information about the filtering of
the General Motors high-frequency data. All numbers are averages
across the 1,255 trading days in our sample from January 3, 2000
through December 31, 2004. \#$r_{ \tau_{i}} \neq 0$ is the daily
amount of tick data left after counting out price repetitions in
consecutive ticks. \#$\Delta
 r_{ \tau_{i}} \neq 0$ also 
removes price reversals.
\end{table}
\vspace*{\fill}

%% file: tables/GM-qu-Sample.tex
\vspace*{\fill}
\begin{table}[H]
\caption{Sample statistics for $RV^{ \Delta}$ and $RRV_{m}^{ \Delta}$.}
\label{GM-qu-Sample.tex}
\begin{tabular*}{\linewidth}{@{~}*{1}{l}*{9}{r}@{~}}
\\[-0.25cm]
\hline
\rule[-3mm]{0mm}{8mm} & Mean & Var. & Skew. & Kurt. & Min. & Max. & & \multicolumn{2}{c}{Correlation} \\
 &  &  &  &  &  &  &  & $RV^{ \Delta}$ & $RRV_{m}^{ \Delta}$\\
$RV^{ \Delta}$ &  7.276 & 47.740 &  3.624 & 25.085 &  0.472 & 79.332 &  &  1.000\\
$RRV_{m}^{ \Delta}$ &  6.212 & 27.693 &  3.056 & 18.417 &  0.496 & 56.256 &  &  0.982 &  1.000\\
\\[-0.35cm]
\hline
\end{tabular*}
\vspace*{0.15cm}

\noindent The table reports sample statistics of
the annualized percentage $RV^{ \Delta}$ and $RRV_{m}^{ \Delta}$
for General Motors during January 3, 2000 up to December 31,
2004. We provide the mean, variance, skewness, kurtosis, minimum
and maximum of the two time series, plus their correlation.\end{table}
\vspace*{\fill}